\documentclass[aps,prl,twocolumn,superscriptaddress,nofootinbib]{revtex4-2}
\usepackage{mathrsfs}
\usepackage{amsmath,amssymb,bm,mathtools}
\usepackage{blindtext}
\usepackage{graphicx}
\graphicspath{ {figures/}{figures/eps/}{figures/pdf/} }
\usepackage{array}
\usepackage{float}
\usepackage[export]{adjustbox}
\usepackage{framed}
\usepackage{physics}
\usepackage{xcolor}
\usepackage[justification=raggedright, font=small]{caption}

\begin{document}

\title{Novel bipartite entanglement in the quantum dimer magnet Yb$_2$Be$_2$SiO$_7$}

\author{A. Brassington} 
\affiliation{Department of Physics and Astronomy, University of Tennessee, Knoxville, TN 37996, USA}

\author{Q. Ma} 
\affiliation{Neutron Scattering Division, Oak Ridge National Laboratory, Oak Ridge, TN 37831, USA}

\author{G. Duan} 
\affiliation {School of Physics and Beijing Key Laboratory of Optoelectronic Functional Materials and Micro-nano Devices, Renmin University of China, Beijing 100872, China}

\author{S. Calder} 
\affiliation{Neutron Scattering Division, Oak Ridge National Laboratory, Oak Ridge, TN 37831, USA}

\author{A.I. Kolesnikov} 
\affiliation{Neutron Scattering Division, Oak Ridge National Laboratory, Oak Ridge, TN 37831, USA}

\author{K.M. Taddei} 
\affiliation{Neutron Scattering Division, Oak Ridge National Laboratory, Oak Ridge, TN 37831, USA}

\author{G. Sala} 
\affiliation{Oak Ridge National Laboratory, Oak Ridge, TN 37831, USA}

\author{E.S. Choi} 
\affiliation{National High Magnetic Field Laboratory and Department of Physics, Florida State University, Tallahassee, Florida 32310, USA}

\author{H. Wang} 
\affiliation {Department of Chemistry, Michigan State
University, East Lansing, Michigan 48824, United States}

\author{W. Xie} 
\affiliation {Department of Chemistry, Michigan State
University, East Lansing, Michigan 48824, United States}

\author{B.A. Frandsen}
\affiliation{Department of Physics and Astronomy, Brigham Young University, Provo, UT 84602, USA}

\author{N. Li}
\affiliation{Anhui Provincial Key Laboratory of Magnetic Functional Materials and Devices, Institutes of Physical Science and Information Technology, Anhui University, Hefei, Anhui 230601, People's Republic of China}

\author{X.F. Sun}
\affiliation{Anhui Provincial Key Laboratory of Magnetic Functional Materials and Devices, Institutes of Physical Science and Information Technology, Anhui University, Hefei, Anhui 230601, People's Republic of China}

\author{C. Liu}
\affiliation {School of Engineering, Dali University, Dali, Yunnan 671003, China}

\author{R. Yu} 
\affiliation {School of Physics and Beijing Key Laboratory of Optoelectronic Functional Materials and Micro-nano Devices, Renmin University of China, Beijing 100872, China}
\affiliation {Key Laboratory of Quantum State Construction and Manipulation (Ministry of Education), Renmin University of China, Beijing 100872, China}

\author{H.D. Zhou} \altaffiliation{\href{mailto:hzhou10@utk.edu}{hzhou10@utk.edu}}
\affiliation{Department of Physics and Astronomy, University of Tennessee, Knoxville, TN 37996, USA}

\author{A.A. Aczel} \altaffiliation{\href{mailto:aczelaa@ornl.gov}{aczelaa@ornl.gov}}
\affiliation{Neutron Scattering Division, Oak Ridge National Laboratory, Oak Ridge, TN 37831, USA}
 
%\date{\today}

\begin{abstract}
The quantum dimer magnet, with antiferromagnetic intradimer and interdimer Heisenberg exchange between spin-1/2 moments, is known to host an $({\ket{\uparrow \downarrow} - \ket{\downarrow \uparrow}})/\sqrt{2}$ singlet ground state when the intradimer exchange is dominant. Rare-earth-based quantum dimer systems with strong spin-orbit coupling offer the opportunity for tuning their magnetic properties by using magnetic anisotropy as a control knob. Here, we present bulk characterization and neutron scattering measurements of the quantum dimer magnet Yb$_2$Be$_2$SiO$_7$. We find that the Yb$^{3+}$ ions can be described by an effective spin-1/2 model at low temperatures and the system does not show signs of magnetic order down to 50~mK. The magnetization, heat capacity, and neutron spectroscopy data can be well-described by an isolated dimer model with highly anisotropic exchange that stabilizes a singlet ground state with a wavefunction $({\ket{\uparrow \uparrow} - \ket{\downarrow \downarrow}})/\sqrt{2}$ or $({\ket{\uparrow \uparrow} + \ket{\downarrow \downarrow}})/\sqrt{2}$. Our results show that strong spin-orbit coupling can induce novel entangled states of matter in quantum dimer magnets. 
\end{abstract}  

\maketitle 

Insulating magnets devoid of ground states with long-range order and instead characterized by strongly-interacting units featuring quantum entanglement \cite{Sachdev_2008,Balents_2010} have attracted intense interest in the field of quantum magnetism. The simplest case consists of a series of dimers with two entangled spins per unit \cite{Bose_2005, Zapf_2014}, which is known as a quantum dimer magnet for spin-1/2. The Shastry-Sutherland lattice (SSL) consists of a plane of orthogonal dimers \cite{Shastry_1981, Liu2014}, and it is now known as one of the canonical two-dimensional models that can host entangled spin states. For isotropic Heisenberg spins on the SSL with a sufficiently small ratio of interdimer-to-intradimer exchange interactions $\alpha =$~$J_2$/$J_1$, the Hamiltonian is exactly solvable and the ground state is predicted to be a direct product of entangled spin-singlet states. The gapped elementary excitations associated with this exotic state are known as triplons \cite{Romhanyi2015,Giamarchi2008}. 

SSL research progressed beyond the initial theoretical studies of the simple toy model with the discovery of SrCu$_2$(BO$_3$)$_2$ \cite{99_kageyama, 99_Kageyama_2}, which consists of weakly-coupled planes of $S = \frac{1}{2}$ Cu$^{2+}$ ions with the desired orthogonal dimer geometry \cite{99_miyahara}. The $\alpha$ ratio is sufficiently small to generate a spin-singlet ground state \cite{99_kageyama, 99_Kageyama_2}, although it is close to the critical value required to produce Neel antiferromagnetic order \cite{99_miyahara, 99_weihong}. This proximity to a quantum phase transition has now been investigated in great detail by both experiment and theory, with a pressure-induced singlet plaquette phase first identified by neutron spectroscopy \cite{Zayed_2017} and multiple theoretical studies predicting quantum spin liquid phases in this regime as well \cite{Wang_2022, Yang_2022, Corboz_2025, Viteritti_2024}. SrCu$_2$(BO$_3$)$_2$ also exhibits striking magnetic-field-induced phenomena, with several magnetization plateaus observed that were first explained by triplon crystallization \cite{02_kodama, 08_sebastian, 12_jaime}. The lower-field plateaus were later argued to arise from triplon bound state crystallization instead \cite{14_corboz}.

More recently, the effect of strong spin-orbit-coupling on the magnetic properties of the SSL has received significant attention. Relevant theoretical work has explored the magnetic properties of the Ising or XXZ SSL Hamiltonian \cite{08_meng, 10_suzuki, 12_dublenych, 14_su, 15_zhang, 24_yadav}, which can be realized for systems with significant magnetic anisotropy. On the experimental side, the $R$B$_4$ ($R =$~rare earth) family members were initially explored as possible anisotropic SSL systems \cite{MICHIMURA_2006, Brunt_2018, Yoshii_2008, Wierschem_2015, Orendac_2021, Matas_2010}. Unfortunately, they are itinerant magnets with long-range Ruderman-Kittel-Kasuya-Yosida (RKKY) exchange interactions mediated by the conduction electrons \cite{14_feng, 18_wierschem, 19_farkasovsky, 19_regeciova, Regeciova_2020}, so their magnetic Hamiltonians are more complex than the simple $J_1$-$J_2$ (or nearest-neighbor and next-nearest-neighbor) SSL model. Two insulating families Ba$R_2$M$X_5$ ($M =$~Zn, Pd, or Pt; $X =$~O or S) \cite{03_wakeshima, 05_ozawa, 08_ozawa, 20_ishii,  Billingsley_2022, Marshall_2023, 23_pasco, Ma_2024} and $R_2$Be$_2Z$O$_7$ ($Z =$~Ge or Si) \cite{Ashtar_2021, Brassington_2024} have now come to the forefront as superior model systems for exploring SSL physics in strongly-anisotropic magnets. A subset of these rare-earth-based materials host effective spin-1/2 degrees of freedom at sufficiently low temperatures, which arise from a well-isolated crystal field ground state doublet.  While ordered ground states have often been identified \cite{05_ozawa, 08_ozawa, Ishii_2021, Billingsley_2022, 23_pasco, 23_song, Brassington_2024_2}, effective spin-1/2 systems can realize more exotic behavior. Both BaCe$_2$ZnS$_5$ \cite{Ma_2024} and Yb$_2$Be$_2$GeO$_7$ \cite{24_pula, 24_liu} are effective spin-1/2 dimer systems with entangled ground states and the latter has emerged as a quantum spin liquid candidate \cite{24_pula}.

In this work, we report comprehensive bulk characterization and neutron scattering measurements of the quantum dimer magnet Yb$_2$Be$_2$SiO$_7$ with SSL geometry. We find that the Yb$^{3+}$ ions have effective spin-1/2 degrees of freedom at sufficiently low temperatures below 10 K, and we observe no evidence for magnetic order down to dilution-refrigerator temperatures of 50~mK. The strong magnetic anisotropy of this system generates nearly Ising moments with a strong tendency to align along the $[001]$-axis. The zero-field neutron spectroscopy and the magnetic-field and temperature-dependence of both the magnetization and specific heat data closely match expectations for a novel entangled dimer state. More specifically, Yb$_2$Be$_2$SiO$_7$ consists of a series of dimers with $S_z \neq 0$ ground state wavefunctions $({\ket{\uparrow \uparrow} - \ket{\downarrow \downarrow}})/\sqrt{2}$ or $({\ket{\uparrow \uparrow} + \ket{\downarrow \downarrow}})/\sqrt{2}$. The anisotropic intradimer exchange induced by the strong spin-orbit-coupling in this system is responsible for stabilizing this exotic state that cannot be realized by the Heisenberg SSL model for spin-1/2 systems. 

\begin{figure}
\scalebox{0.455}{\includegraphics{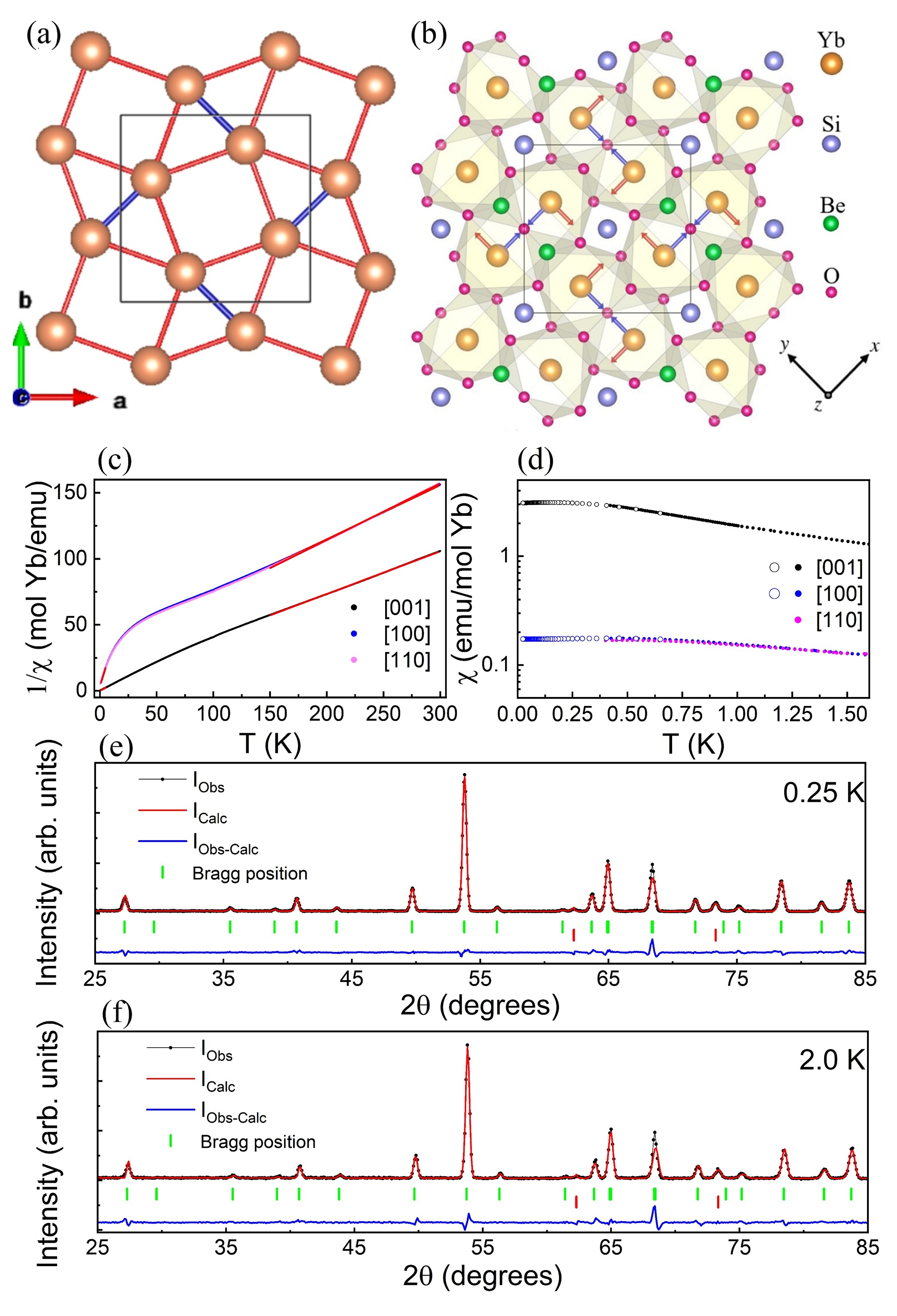}}
\caption{{\bf Crystal structure and magnetic susceptibility.} (a) The arrangement of Yb$^{3+}$ ions in Yb$_2$Be$_2$SiO$_7$ forming a SSL as viewed along the crystallographic [001]-axis. The intradimer and interdimer bonds $J_1$ and $J_2$ are shown in blue and red respectively. (b) The full crystal structure of Yb$_2$Be$_2$SiO$_7$ viewed along the same axis. The four Yb$^{3+}$ ions in the chemical unit cell have diagonal g-tensors with different local axes, as explained in the main text. (c) The inverse susceptibility measured in a 0.1 T field applied along three high-symmetry crystallographic directions. The Curie-Weiss fits in both the high and low-temperature regimes are shown in red. (d) The low-$T$ susceptibility data plotted along three high-symmetry crystallographic directions. Closed circles and open circles represent DC and AC susceptibility data respectively. (e-f) Neutron powder diffraction patterns collected at 0.25 K and 2 K, respectively. Multi-phase Rietveld refinements with the known Yb$_2$Be$_2$SiO$_7$ crystal structure (green ticks) and Al from the sample can (red ticks) explain the data well. No additional peaks associated with long-range order are observed.}
\label{fig1}
\end{figure}

\section{Results}

\subsection{Crystal structure and magnetic susceptibility}

Yb$_2$Be$_2$SiO$_7$ crystallizes into the tetragonal space group \textit{P-42$_1$m} (113) with room temperature lattice constants $a =$~7.207(1)~\AA~and $c =$~4.719(1)~\AA~determined from single crystal x-ray diffraction (see Supplementary Discussion Section I and Supplementary Table S1 and S2 for more details). The eight-fold coordinated Yb$^{3+}$ ions form an SSL lattice in the $ab$-plane with intradimer and interdimer distances of 3.239(1)~\AA~and 3.836(1)~\AA~respectively as shown in Fig.~\ref{fig1}(a). Adjacent Yb$^{3+}$ planes are separated by 4.719(2)~\AA. A view of the crystal structure along the $c$-axis with all the atoms shown is depicted in Fig.~\ref{fig1}(b).

\begin{figure*}
\scalebox{0.53}{\includegraphics{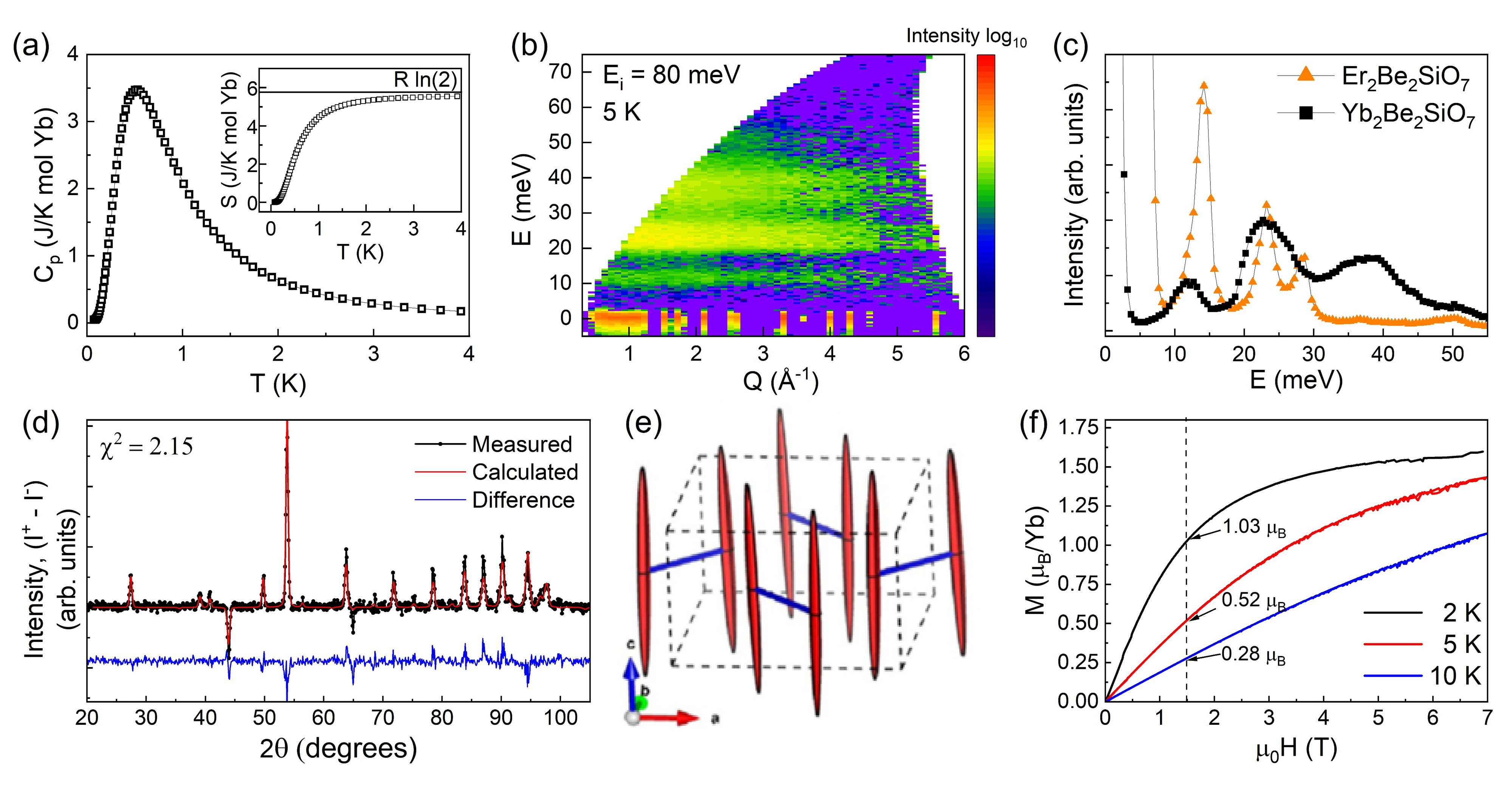}}
\caption{{\bf Yb$^{3+}$ single ion ground state.} (a) Zero-field heat capacity data reveals a broad Schottky anomaly centered near 0.5~K. The inset shows the magnetic entropy over the same temperature range, which approaches $Rln(2)$ at 4~K. (b) Color contour plot of the scattering intensity for the Yb$^{3+}$ ions in Yb$_2$Be$_2$SiO$_7$ as a function of momentum and energy transfer at 5~K from SEQUOIA with $E_i =$~80 meV. This difference map was obtained by subtracting scaled Lu$_2$Be$_2$SiO$_7$ data to account for the variation in the Yb and Lu neutron scattering cross sections. (c) Constant-$Q$ cuts ($Q$ integration range $[0.25, 5]$~\AA$^{-1}$) of the same Yb$_2$Be$_2$SiO$_7$ data (with no Lu subtraction) and the equivalent data for the magnetic Er \cite{Brassington_2024_2} analog. (d) The difference pattern, I$^+$ - I$^-$, obtained from the pNPD measurements at 5 K. The Rietveld refinement is superimposed on the data and the fit residual is shown below it. (e) The magnetization ellipsoids obtained from the pNPD analysis at 5~K. The Yb$^{3+}$ anisotropy is nearly Ising with a strong tendency for the moments to align close to the crystallographic $c$-axis. (f) Magnetization vs applied magnetic field for polycrystalline Yb$_2$Be$_2$SiO$_7$ at 2~K, 5~K, and 10 K. The measured values at 1.5~T are labeled on the panel.}
\label{fig2}
\end{figure*}

The inverse magnetic susceptibility of single crystalline Yb$_2$Be$_2$SiO$_7$, plotted as field/magnetization $H/M$, is presented in Fig.~\ref{fig1}(c) between 0.4~K and 300~K. The data was collected in a small magnetic field of 0.1~T applied along the three high-symmetry directions $[001]$, $[110]$, and $[100]$. The same data is shown in the low-$T$ range between 0.4~K and 1.6~K in Fig.~\ref{fig1}(d), with the real component of the AC susceptibility data from 0.05-0.65~K also included for two field orientations. There are no clear signatures of long-range magnetic order in this data. Curie-Weiss fits were performed over both high-$T$ (150~K - 300~K) and low-$T$ (1~K - 5~K) ranges. The effective moment values from the high-$T$ fits are broadly consistent with expectations for free Yb$^{3+}$ ions. The low-$T$ fits return effective moments and Curie-Weiss temperatures of 4.22(5)~$\mu_B$ and -0.10(2)~K for $\vec{H} \parallel [001]$, 1.71(1)~$\mu_B$ and -1.32(1)~K for $\vec{H} \parallel [110]$, and 1.67(1)~$\mu_B$ and -1.21(1)~K for $\vec{H} \parallel [100]$. The comparatively large effective moment for $\vec{H} \parallel [001]$ indicates that the moments have a strong preference to align with the crystallographic $[001]$-axis at low-$T$. The small negative Curie-Weiss temperatures are indicative of weak antiferromagnetic interactions in the system. 

Unpolarized neutron powder diffraction (NPD) patterns collected at 0.25~K and 2~K are presented in Fig.~\ref{fig1}(e) and (f) respectively. Both patterns are well-described by nuclear Bragg peaks associated with the Yb$_2$Be$_2$SiO$_7$ crystal structure and the Al sample can. The expected positions of these peaks are indicated by the green and red ticks respectively. While Rietveld refinements of the data using FULLPROF \cite{RODRIGUEZCARVAJAL_1993} reveal no evidence for oxygen off-stoichiometry in this sample, they do identify $\sim$4.4\% Be/Si site mixing. Most importantly, there are no additional Bragg peaks or enhanced peak intensities that can be attributed to long-range magnetic order in Yb$_2$Be$_2$SiO$_7$, even down to 0.25~K. Additional details of the 0.25~K refinement are presented in Supplementary Table S3.

\subsection{Yb$^{3+}$ single ion ground state}

Zero-field specific heat data measured with a single crystal and the corresponding magnetic entropy extracted by integrating $C_p/T$ as a function of temperature are shown in Fig.~\ref{fig2}(a) and its inset respectively. There are no typical features of magnetic ordering in the heat capacity data. Instead, it is dominated by a broad peak centered near 0.5~K, while the magnetic entropy approaches $Rln(2)$ at 4~K. The entropy result validates the use of an effective spin-1/2 model for the Yb$^{3+}$ ions at low-$T$ and indicates that the broad peak does not have a crystal field origin. Neutron powder spectroscopy, with the scattering intensity plotted as a function of momentum transfer $Q$ and energy transfer $E$ in Fig.~\ref{fig2}(b), was used to measure the Yb$^{3+}$ crystal field spectrum. The magnetic scattering contribution was isolated by subtracting the scattering from the non-magnetic analog Lu$_2$Be$_2$SiO$_7$ measured under the same experimental conditions and scaled appropriately to account for the scattering cross-section difference between Yb and Lu. There is no evidence for low-lying crystal field levels in this data. Instead, there are three excitations centered about 11, 23, and 36~meV as shown in Fig.~\ref{fig2}(c) that likely correspond to the three excited crystal field doublets expected for $J =$~7/2 Kramers Yb$^{3+}$ ions in a low-symmetry ligand environment. Interestingly, these three excitations are much broader than the instrument energy resolution, which is readily apparent when comparing this data to previous results from Er$_2$Be$_2$SiO$_7$ \cite{Brassington_2024_2} with nearly resolution-limited crystal field levels. This broadening may be a consequence of the Be/Si site mixing that we identified in the Rietveld refinements of the HB-2A data. Interestingly, we confirmed that there is no evidence for Be/Si site mixing in the previously reported HB-2A data for Er$_2$Be$_2$SiO$_7$ \cite{Brassington_2024_2}. We also collected Pair-Distribution-Function (PDF) data using the time-of-flight powder diffractometer NOMAD to investigate possible local structural distortions in Yb$_2$Be$_2$SiO$_7$ (see Supplementary Discussion Section I and Supplementary Fig.~S1), but the PDF data is explained well by the global structure. 

The Yb$^{3+}$ ions have a monoclinic point group symmetry of C$_s$ with a single mirror plane symmetry element. The crystal field environment of the four Yb$^{3+}$ ions in the chemical unit cell is different, with the $y$-axis of the local coordinate system indicated by the orange arrows in Fig.~\ref{fig1}(b). The local $y$-axes are parallel to the mirror-plane normals and perpendicular to the intradimer bonds. One principal g-tensor direction lies along the local $y$-axis. The other two principal g-tensor directions are not constrained by symmetry and make an angle of $\beta$ with the crystallographic $[001]$ and $[110]$ (or $[1\bar{1}0]$ directions). The local coordinate system sketched in Fig.~\ref{fig1}(b) applies when $\beta =$~0. Under this assumption, the blue arrows in this same panel denote the local $x$-axes, which are parallel to the dimer bonds.

Due to the low point symmetry of the Yb$^{3+}$ ions and the lack of sufficient crystal field observables in the neutron spectroscopy data, it is not feasible to obtain a unique set of crystal field parameters for Yb$_2$Be$_2$SiO$_7$. Half-polarized neutron powder diffraction (pNPD) offers a powerful alternative to probe the single ion anisotropy of this system via the local site susceptibility approach \cite{Kibalin_2019}. pNPD is performed in the paramagnetic region and therefore the lack of long-range order does not influence the application of this technique. Spin-up and spin-down neutron diffraction patterns with intensities of I$^+$ and I$^-$ are measured separately, and then Rietveld refinements of the sum and difference patterns (given by I$^+$ + I$^-$ and I$^+$ - I$^-$ respectively) are performed using the software CrysPy \cite{Kibalin_2019}. The Yb$^{3+}$ local site susceptibility tensor, which provides information on the local anisotropy and principal $g$-tensor directions and their relative magnitudes, can be extracted from the difference pattern. 

The difference pattern I$^+$ - I$^-$ measured at 5~K in an applied magnetic field of 1.5~T is shown in Fig.~\ref{fig2}(d) and the sum pattern from the same dataset is shown in Supplementary Fig.~S2. In each case, the Rietveld refinement result is superimposed on the data and the fit residual is plotted below the data. The best-fit result of the difference pattern yields a small $\beta$ value of 1.52(3)$^\circ$, which means that the two principal $g$-tensor directions not constrained by symmetry are nearly aligned with the crystallographic $[001]$-axis and the dimer bond direction ($[110]$ and $[1\bar{1}0]$ for the two dimer sublattices) as illustrated in Fig.~\ref{fig1}(b). The magnetization ellipsoids' principal axes have magnitudes of 2.32 $\mu_B$/T, 0.14 $\mu_B$/T, and 0.10 $\mu_B$/T, with the largest value for the pseudo-$[001]$-axis direction. These results are indicative of an Ising-like anisotropy with small deviations away from the $[001]$-axis and they are consistent with our Curie-Weiss fitting results described above. The analysis of the 10~K difference pattern measured in the same 1.5~T magnetic field produces similar results. The magnetization ellipsoids calculated from the local site susceptibility analysis are presented in Fig.~\ref{fig2}(e). CrysPy can also be used to calculate the powder-averaged magnetization from the local-site susceptibility tensor to facilitate a direct comparison to the experimental data as a sanity check. The powder-averaged magnetization at selected temperatures for Yb$_2$Be$_2$SiO$_7$ is presented in Fig.~\ref{fig2}(f) with the measured 1.5~T values shown on the panel. The 5~K and 10~K values calculated from the refined local site susceptibility tensors are 0.53 $\mu_{B}$/Yb and 0.29 $\mu_{B}$/Yb~respectively, which are nearly the same values from our powder-averaged magnetization measurements. 

\begin{figure}
\scalebox{0.19}{\includegraphics{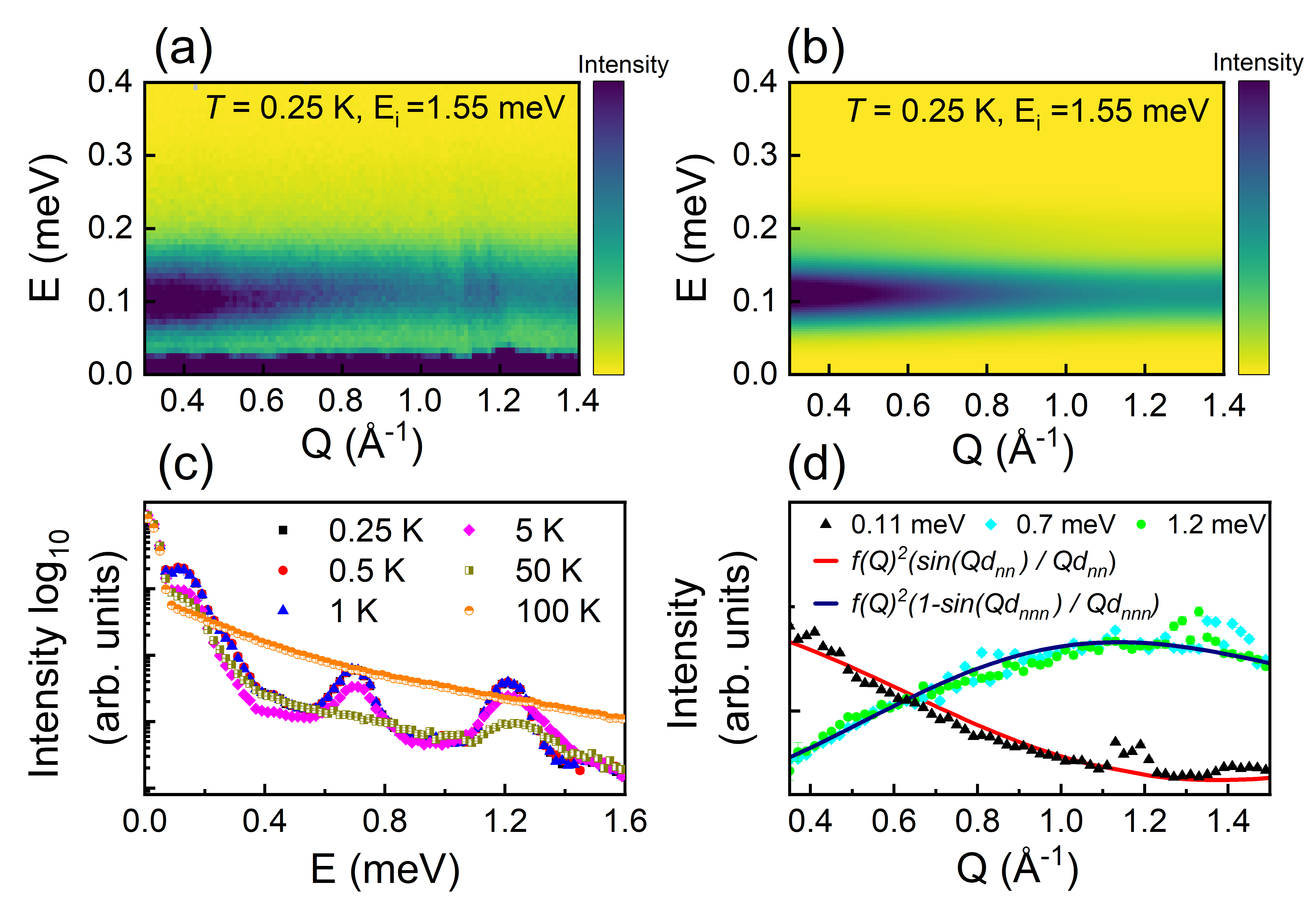}}
\caption{{\bf Dynamical structure factor.} (a) Color contour plot of the scattering intensity as a function of momentum and energy transfer at 0.25~K from CNCS with $E_i =$~1.55~meV. A weakly-dispersive mode with a magnetic origin is centered at 0.11~meV. (b) Color contour plot of the simulated scattering intensity using an isolated dimer model with the anisotropic intradimer exchange parameters provided in the main text. (c) Constant-$Q$ cuts of the scattering intensity ($Q$ integration range $[0.5, 1.8]$~\AA$^{-1}$) with $E_i =$~2.49~meV at selected temperatures. Aside from the main excitation band centered at 0.11~meV with a shoulder at 0.19~meV, there are two higher excitation bands centered at 0.7~meV and 1.2~meV respectively. (d) Constant-$E$ cuts centered about three different mode positions, with an energy integration range $\pm$~0.05~meV for the lowest mode ($E_i =$~1.55~meV data) and an integration range of $\pm$~0.1~meV for the two higher modes ($E_i =$~2.49~meV data). The intensities of the two higher-energy cuts have been rescaled to facilitate a straightforward comparison of their $Q$-dependence, which is strikingly different from the lower excitation band. Simulations based on two different types of single dimer ground states are superimposed on the data and describe it well.} 
\label{fig3}
\end{figure}

\subsection{Isolated dimer model}

Low-energy neutron powder spectroscopy data were used to investigate the magnetic excitations associated with the collective ground state of Yb$_2$Be$_2$SiO$_7$. A color contour plot of the scattering intensity as a function of $Q$ and $E$ with $T =$~0.25~K measured on CNCS with $E_i =$~1.55~meV is presented in Fig.~\ref{fig3}(a). A simulated scattering pattern based on an isolated dimer model that will be discussed below with Gaussian peak widths (full-width half maximums) that match the expected instrumental energy resolution at the mode positions (i.e. 0.035~meV at 0.11~meV and 0.033~meV at 0.19~meV) is depicted in Fig.~\ref{fig3}(b). A low-energy band of excitations with minimal dispersion centered at 0.11~meV and an intensity maximum as $Q \rightarrow$~0 is visible in the data. There is also a second excitation at 0.19~meV that manifests as a shoulder in the data. Both mode energies were determined by fitting the data to a sum of Gaussian functions as shown in Supplementary Fig.~S3 and summarized in Supplementary Table~S4. Constant-$Q$ cuts at selected temperatures are presented in Supplementary Fig.~S4 for the $E_i =$~1.55~meV dataset and over a wider energy range in Fig.~\ref{fig3}(c) for the $E_i =$~\textcolor{red}{2.49}~meV dataset. The logarithmic intensity scales help to reveal two higher-energy excitation bands centered at 0.7~meV and 1.2~meV. The $Q$-dependence of the three distinct modes is shown in Fig.~\ref{fig3}(d). Both the $Q$ and $T$-dependence of these modes indicate that they are magnetic excitations and their nearly-dispersionless nature suggests that they are associated with isolated or weakly-interacting spin dimers. 

Since effective spin-1/2 dimer models with strong intradimer exchange anisotropy can have a maximum of three single-dimer excitations, it is clear that all four modes (at 0.11, 0.19, 0.7 and 1.2 meV) observed here do not have the same origin. Zero-field heat capacity only depends on the eigenvalues of these models, so we compared our data shown in Fig.~\ref{fig2}(a) to simulations assuming various energy-level schemes (see Supplementary Fig.~S5). The best simulation of the zero-field heat capacity data corresponds to an isolated dimer model with a doubly-degenerate mode at 0.11~meV and a non-degenerate mode at 0.19~meV. This implies that the two higher-energy modes have a different origin. In fact, as shown in Fig.~\ref{fig3}(d) their $Q$-dependence is well-described by the Yb$^{3+}$ magnetic form factor squared multiplied by the known structure factor of a singlet-triplet transition for an isolated Heisenberg dimer model \cite{05_haraldsen} with the square plaquette distance $d_{nnn} =$~3.84~\AA. Since these higher-energy modes have a negligible contribution to the specific heat, this implies that only a small fraction of Yb$^{3+}$ ions form longer-range dimers in Yb$_2$Be$_2$SiO$_7$, possibly due to the known Be/Si site mixing in this system. We also compared the $Q$-dependence of these modes to expectations for the lowest-energy excitations of rectangular tetramers \cite{05_haraldsen}, but this model could not account for the $Q$-position of their maximum intensities. Finally, we note that higher-energy excitations in SrCu$_2$(BO$_3$)$_2$ have been attributed to singlet bound states consisting of contributions from multiple dimer units \cite{McClarty_2017, Wulferding_2021, Fogh_2024} and a similar alternative origin for the higher-energy modes cannot be ruled out in the present case. 

With the energy level scheme for a possible isolated dimer model comprised of most Yb$^{3+}$ ions in the system established, we performed a quantitative analysis of the low-energy neutron spectroscopy data and field-dependent heat capacity and magnetization data. We started by constructing a minimal model based on symmetry analysis \cite{Ma_2024,Liu_2024}. The model Hamiltonian describes a collection of isolated dimers with anisotropic intradimer interactions and Zeeman coupling:
\begin{equation}\label{Eq:Ham}
\mathcal{H} = \sum_{<i,j>l\alpha} S^{\alpha}_{il} J^l_{\alpha\alpha} S^{\alpha}_{jl} - \mu_B \sum_{il\alpha\beta} H^{\alpha} g^{l}_{\alpha\beta} S^{\beta}_{il}
\end{equation}
%$$ 
%$$ + \sum_{<i,j>} \mathbf{S}^{B}_{i} \cdot \mathbf{J}^{B} \cdot \mathbf{S}^{B}_{j} - \mu_0\mu_B  \sum_{<j>} \mathbf{H} \cdot  \mathbf{g}^{B} \cdot \mathbf{S}^{B}_{j}$$
where the superscript $l$ denotes dimers in sub-lattice A/B, $S^{\alpha}_i$ refers to the $\alpha (=x, y, z)$ component of the effective spin 1/2 operator with $x =[1,1,0]$, $y =[-1,1,0]$ and $z = [0,0,1]$, $J^{l}_{\alpha\alpha}$ represents one component of the intradimer exchange tensor for dimers in each sub-lattice, $H^{\alpha}$ is the $\alpha$ component of the applied magnetic field, and $g^{l}_{\alpha\beta}$ represents one component of the $g$-tensor. Our model is defined in global crystallographic coordinates and it is equivalent to the one defined within the local basis after proper rotations. Constrained by the time-reversal and mirror symmetries, this Hamiltonian takes a diagonal form with an $XYZ$ type of spin anisotropy in the reference frame defined above, and the couplings on the two sub-lattices are related by a $90^{\circ} $ rotation about the crystallographic $[001]$-axis so that $J^{A}_{xx/yy} =J^{B}_{yy/xx}$ and $J^{A}_{zz} =J^{B}_{zz}$. 

\begin{figure}
\scalebox{0.23}{\includegraphics{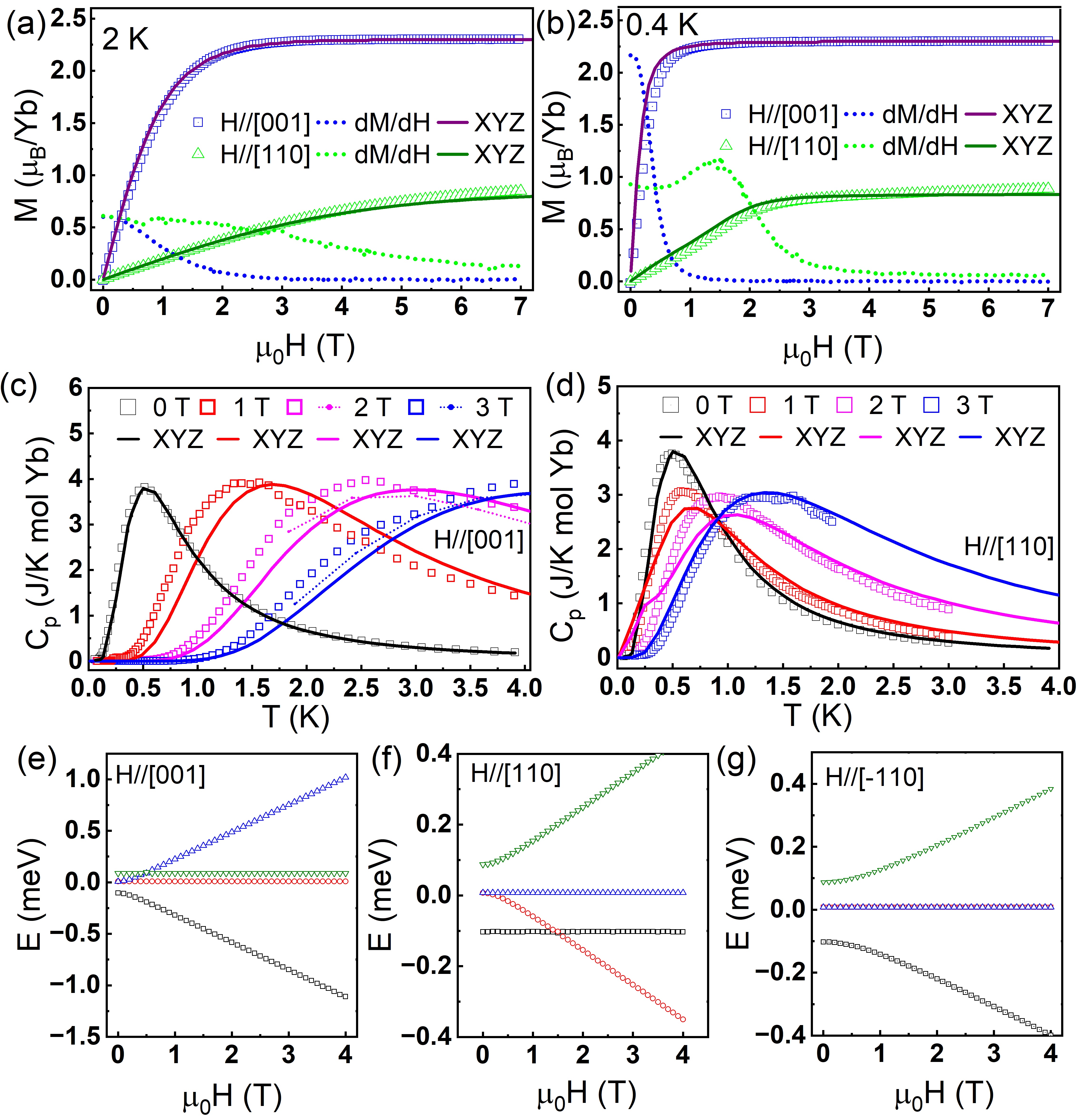}}
\caption{{\bf Isolated dimer modeling.} (a) Magnetization vs field along two high-symmetry crystallographic directions at 2~K. (b) Similar data collected at 0.4~K. (c) Heat capacity vs magnetic field applied along the $[001]$ direction. Experimental data are shown as open symbols while simulated data from our XYZ anisotropic exchange model are shown as solid curves. (d) Heat capacity data vs magnetic field applied along the $[110]$ direction with simulated data from the same model superimposed on it. (e-g) Simulated isolated dimer energy levels vs applied field along different crystallographic directions for one dimer sublattice based on the XYZ anisotropic exchange model. There is a level crossing between 1.5~T and 2~T for one dimer sublattice when $\vec{H} \parallel [110]$ or for the other dimer sublattice when $\vec{H} \parallel [\bar{1}10]$.}
\label{fig4}
\end{figure}

Although off-diagonal $g$-tensor components are allowed by symmetry, our pNPD results described above suggest they are much smaller than the diagonal ones. We hence take the approximation that the g-tensor follows the same symmetry as the exchange tensor. With the principal $g$-tensor directions now fixed along high-symmetry crystallographic directions, we estimated the $g$-tensor values by using the 2~K magnetization data presented in Fig.~\ref{fig4}(a) and additional $\vec{H} \parallel [100]$ magnetization data (not shown). After subtracting off a linear van-Vleck contribution in the highest-field regime, we obtain $g^{A}_{xx}=1.64~\mu_B$, $g^{A}_{yy}=1.71~\mu_B$, and $g^{A}_{zz}=4.6~\mu_B$. 

The energy level scheme established from the neutron spectroscopy and zero-field heat capacity data described above is consistent with 12 different intradimer exchange matrices, which we determined using Supplementary Eq.~S2. We then performed exact diagonalization of these 12 Hamiltonians in zero field to find six solutions with $S_z =$~0 ground state wavefunctions and six solutions with $S_z \neq$~0 ground state wavefunctions. We calculated the dimer structure factors for the three excitations of each model using Supplementary Eq.~S4, with the results for three representative models shown in Supplementary Fig.~S6. The two types of solutions are characterized by intense dimer excitations with very different $Q$-dependences. In the regime $Q \le$~1.6~\AA$^{-1}$, the intense excitation for the $S_z =$~0 solutions is described by the structure factor,
\begin{equation}
S(Q) = A\left(1-\frac{sin(Qd)}{Qd}\right)
\end{equation}
where $A$ is a constant and $d$ is the intradimer distance. In the same $Q$-regime, the intense excitation for the $S_z \neq$~0 solutions is described approximately by the structure factor
\begin{equation}
S(Q) \approx A\frac{sin(Qd)}{Qd}.
\end{equation}
To narrow down possible intradimer exchange matrices, we examined the $Q$-dependence of the constant $E$-cut centered at the intense 0.11~meV mode with an integration range of $\pm$0.05~meV shown in Fig.~\ref{fig3}(d). The data fits well to the expected structure factor for the intense mode of the six $S_z \neq$~0 solutions. Two of these options can be ruled out because their most intense mode is located at 0.19~meV, which is inconsistent with the neutron spectroscopy data. This leaves four viable solutions and the corresponding simulations for them are presented in Supplementary Fig.~S7. 

All four remaining candidate models have single-dimer ground states of the form $({\ket{\uparrow \uparrow} + \ket{\downarrow \downarrow}})/\sqrt{2}$ or $({\ket{\uparrow \uparrow} - \ket{\downarrow \downarrow}})/\sqrt{2}$ with a dominant contribution from the other $S_z \neq 0$ wavefunction to the doubly-degenerate first excited state. We note that these two dimer ground states are qualitatively consistent with the low-$T$ magnetic susceptibility data presented in Fig.~\ref{fig1}(d), which should level off at finite values for all field orientations as $T \rightarrow$~0 rather than drop to zero as expected for a Heisenberg dimer model. Two of the solutions show superior agreement with the bulk characterization data. Although the $Q$-dependence of the weak mode intensities from neutron powder spectroscopy is not exactly the same for these two models, our data is not sufficient for differentiating between them conclusively. The exchange parameters for one of these models are $J_{xx} =~0.19$~meV, $J_{yy} = -0.03$~meV, and $J_{zz} = -0.19$~meV and the exchange parameters for the second model are $J_{xx} =~-0.03$~meV, $J_{yy} = 0.19$~meV, and $J_{zz} = -0.19$~meV.

Using the Hamiltonian parameters provided above for the first model, the neutron spectroscopy simulation is shown in Fig.~\ref{fig3}(b) and the calculated single crystal magnetization and heat capacity data are superimposed on the corresponding experimental datasets in Fig.~\ref{fig4}(a-d). The agreement between experiment and theory is excellent, with the small differences between the $\vec{H} \parallel [001]$ heat capacity data and the simulation arising from modest sample misalignment that was confirmed by re-measuring the 2~T and 3~T data with coarse temperature steps for $T \ge$~1.8~K. This simulation yields an effective $S = 1$ dimer state $({\ket{\uparrow \uparrow} - \ket{\downarrow \downarrow}})/\sqrt{2}$ in zero field, which is stabilized by the large ferromagnetic exchange interaction $J_{zz}$ along the quantization axis. The excited states are doubly degenerate $({\ket{\uparrow \uparrow} + \ket{\downarrow \downarrow}})/\sqrt{2}$ and $({\ket{\uparrow \downarrow} - \ket{\downarrow \uparrow}})/\sqrt{2}$ at $E=0.11$ meV and $({\ket{\uparrow \downarrow} + \ket{\downarrow \uparrow}})/\sqrt{2}$ at $E=0.19$ meV, respectively. The second model has a similar eigenvector scheme for its modes, although the two $S_z \neq$~0 modes swap their eigenvalues so the ground state is $({\ket{\uparrow \uparrow} + \ket{\downarrow \downarrow}})/\sqrt{2}$ instead. Both isolated dimer models account for the major experimental features well, including the $Q$ and $E$-dependence of the dynamical structure factor for the dimer excitations and the magnetic-field-dependence of the magnetization and specific heat. These results confirm that the interdimer interactions are significantly weaker than the intradimer interactions in Yb$_2$Be$_2$SiO$_7$. 

Beyond the qualitative signatures of the $S_z \neq 0$ dimer ground state in neutron spectroscopy, we also observe an anisotropic magnetization process that is not expected for systems with $S_z =$~0 dimer ground states. In particular, no low-field magnetization plateaus are observed at 0.4~K when the magnetic field is applied along any high-symmetry crystallographic direction, which is in sharp contrast to expectations for $S_z =$~0 dimer ground states as shown in Supplementary Fig.~S6. Instead, applying a magnetic field along the $[001]$ direction continuously rotates the $(\ket{\uparrow \uparrow} - \ket{\downarrow \downarrow})/\sqrt{2}$ ground state without closing the excitation gap so that the system is immediately magnetized and evolves to the fully-polarized state adiabatically without a transition as shown in the Fig.~\ref{fig4}(b) data and illustrated by the energy level vs field diagram presented in Fig.~\ref{fig4}(e). On the other hand, applying the field along the $[110]$ direction causes a level crossing for one dimer sublattice as shown in Fig.~\ref{fig4}(f), which corresponds to a quantum phase transition to an intermediate state as indicated by the peak in $dM/dH$ shown in Fig.~\ref{fig4}(b). The other dimer sublattice has a similar level crossing when $\vec{H} \parallel [\bar{1}10]$. It is worth noting that the largest discrepancy in the $\vec{H} \parallel$~[110] bulk characterization data and the simulations is observed in the fixed-field heat capacity measurements shown in Fig.~\ref{fig4}(d) near the expected critical field for the level crossing. This is likely due to neglecting the appropriate sub-leading interaction in the simulations, which should have the largest effect on the magnetic properties of the system in this regime \cite{Ma_2024}. Indeed, neutron powder spectroscopy also suggests that a small additional term should be added to the Hamiltonian to capture the increased bandwidth of the 0.11~meV mode observed in the data, but its accurate determination without single-crystal neutron spectroscopy data is challenging. The nature of this term is not obvious either, as the expected sub-1~K energy scale means it could arise from interdimer exchange or a dipolar interaction. While additional theoretical work on candidate magnetic Hamiltonians for this system and its isostructural family members could help to address this issue, neglecting this extra term does not affect the novel zero-field magnetic ground state for Yb$_2$Be$_2$SiO$_7$ established here.

\section{Discussion}

Quantum dimer magnets with pure spin-1/2 degrees of freedom have been studied extensively \cite{Giamarchi2008, Zapf_2014}. Antiferromagnetic Heisenberg exchange at the intradimer level stabilizes the entangled singlet ground state $({\ket{\uparrow \downarrow} - \ket{\downarrow \uparrow}})/\sqrt{2}$ in zero field with a spin gap to a triplet excited state. Key experimental signatures of this exotic ground state are an isotropic decrease in the magnetic susceptibility with decreasing temperature and zero magnetization (i.e. a magnetization plateau) at sufficiently low temperatures and applied fields, along with a powder-averaged dynamical structure factor for the triplet excitation that satisfies Eq.~2 exactly \cite{05_haraldsen}. Quantum dimer magnets with effective spin-1/2 degrees of freedom can exhibit very different behavior due to their realization of anisotropic exchange Hamiltonians. The type of magnetic ion and its point symmetry, along with the composition of the crystal-field ground-state wavefunctions, can give rise to intradimer exchange matrices with different symmetries including Ising, XY, XXZ, or even XYZ. The quasi-orthorhombic point symmetry for the Yb$^{3+}$ ions in Yb$_2$Be$_2$SiO$_7$ generates the XYZ dimer model, which was only recently explored by theorists \cite{24_liu}. Their most important insight was that the ground state of the XYZ model can be any one of the four possible dimer states depending on the sign and relative strength of the three anisotropic exchange matrix terms. One can distinguish between the four dimer ground states via qualitatively different magnetic susceptibilities, magnetizations, and powder-averaged dynamical structure factors, as shown in Supplementary Fig.~S6. We have measured all three quantities for Yb$_2$Be$_2$SiO$_7$ and found that their behavior matches expectations for the novel $({\ket{\uparrow \uparrow} - \ket{\downarrow \downarrow}})/\sqrt{2}$ or $({\ket{\uparrow \uparrow} + \ket{\downarrow \downarrow}})/\sqrt{2}$ ground state.

Our results have broad implications for other quantum dimer magnets with effective spin-1/2 degrees of freedom and show that anisotropic intradimer exchange should always be considered when trying to understand the magnetic properties of these materials. Yb$_2$Si$_2$O$_7$ was the first insulating rare-earth-based quantum dimer magnet discovered with effective spin-1/2 moments \cite{19_hester}. While the bulk characterization and neutron scattering results are broadly consistent with a $({\ket{\uparrow \downarrow} - \ket{\downarrow \uparrow}})/\sqrt{2}$ zero-field ground state, the magnetic Hamiltonian is still unknown. An interacting version of the XYZ dimer model discussed here may be appropriate due to the low-point symmetry of the Yb ions in that system. Anisotropic intradimer exchange should also be important for the new insulating SSL families Ba$R_2$M$X_5$ ($R =$~rare earth; $M =$~Zn, Pd, or Pt; $X =$~O or S) and $R_2$Be$_2Z$O$_7$ ($Z =$~Ge or Si). In fact, it has been shown to stabilize the same $({\ket{\uparrow \uparrow} - \ket{\downarrow \downarrow}})/\sqrt{2}$ dimer ground state in the effective spin-1/2 system BaCe$_2$ZnS$_5$ \cite{Ma_2024} that we have identified as one possibility for Yb$_2$Be$_2$SiO$_7$ here. Despite this progress, direct links between models with anisotropic intradimer exchange and the measured magnetic properties of these materials have remained scarce. There is now ample materials' phase space to explore this topic in more detail.

In summary, we have characterized the low-temperature magnetic properties of the quantum dimer magnet Yb$_2$Be$_2$SiO$_7$ using bulk characterization and neutron scattering techniques. We find that the effective spin-1/2 Yb$^{3+}$ moments form a novel, bipartite entangled state arising from dominant, anisotropic intradimer exchange. The magnetic Hamiltonian for Yb$_2$Be$_2$SiO$_7$ gives rise to a dimer ground state with $S_z \neq 0$ ground state wavefunctions instead of the $({\ket{\uparrow \downarrow} - \ket{\downarrow \uparrow}})/\sqrt{2}$ state commonly realized in Heisenberg dimer systems based on spin-1/2 moments. A significant amount of Be/Si site mixing in Yb$_2$Be$_2$SiO$_7$ may give rise to the broad crystal field levels and possibly the two higher-energy dimer excitations. Our work shows that quantum dimer magnets with strong spin-orbit coupling are promising playgrounds for identifying novel entangled states of matter in quantum materials.  

\section{Methods}

\subsection{Sample preparation}

Polycrystalline samples of Yb$_2$Be$_2$SiO$_7$ were prepared from stoichiometric amounts of Yb$_2$O$_3$, BeO, and SiO$_2$ as detailed in \cite{Brassington_2024}. Small single crystals were grown using the floating zone melt method \cite{Rey-Garcia_2021, Koohpayeh_2008} and oriented via Laue back-diffraction \cite{Greilinger_2015,Warren_1941}. Detailed structure refinements of single-crystal x-ray diffraction data (XRD) agree well with the previously-reported structure obtained from powder XRD \cite{Brassington_2024} and the results are provided in Supplementary Table S1 and Supplementary Table S2.

\subsection{Bulk characterization}

Polycrystalline and single crystal DC magnetization measurements as a function of magnetic field and temperature were performed using a MPMS3 SQUID magnetometer (Quantum Design) equipped with a He-3 insert. For the single crystal measurements, the magnetic fields were applied along high-symmetry crystallographic directions. Specific heat measurements using single crystals were performed with a Quantum Design Physical Property Measurement System (PPMS) equipped with a dilution refrigerator insert. The AC susceptibility measurements were conducted on single crystals with a voltage-controlled current source (Stanford Research, CS580) and lock-in amplifier (Stanford Research, SR830). The phase of the lock-in amplifier was set to measure the first harmonic signal. The RMS amplitude of the AC excitation field was set to be 0.6 Oe with the frequency fixed to be 200 Hz. The measurements were performed in the SCM1 Dilution Refrigerator of the National High Magnetic Field Laboratory, Tallahassee.

\subsection{Neutron scattering}

Neutron powder diffraction (NPD) was performed using the high-resolution HB-2A powder diffractometer \cite{Garlea_2010,Calder_2018} at the High Flux Isotope Reactor (HFIR) of Oak Ridge National Laboratory (ORNL). Unpolarized measurements were conducted with a $\sim$4~g polycrystalline sample sealed inside an aluminum can with 1 atm of helium exchange gas. Diffraction patterns were collected at 250 mK and 2 K with a wavelength of 2.41~\AA~and a collimation of open-open-12’. 

A half-polarized neutron powder diffraction (pNPD) experiment was performed using the same HB-2A collimation settings and a V-cavity to generate the polarized beam with a $\sim$13~g polycrystalline sample pressed into pellets. The sample was loaded in a larger Al can than the one used for the unpolarized measurements. The polarization state was controlled using a Mezei flipper. Further details of the HB-2A pNPD experimental set-up can be found in \cite{Baral_2025}.  Spin-up and spin-down diffraction patterns, with intensities denoted by I$^+$ and I$^-$ respectively, were collected in a vertical magnetic field of 1.5~T at temperatures of 5 and 10~K using a wavelength of 2.41~\AA. 

Time-of-flight neutron powder diffraction data were collected at 100~K on NOMAD \cite{Calder_2018} at the Spallation Neutron Source (SNS) of ORNL using $\sim$4~g of powder. Similar measurements were carried out on $\sim$4.5~g of crushed single crystals and $\sim$2.3~g of polycrystalline Er$_2$Be$_2$SiO$_7$ \cite{Brassington_2024_2}.

Neutron powder spectroscopy data was obtained from the direct-geometry time-of-flight instrument SEQUOIA \cite{Granroth_2010} at the SNS using $\sim$2.5~g of similar Yb$_2$Be$_2$SiO$_7$ powder. The same amount of a non-magnetic reference sample Lu$_2$Be$_2$SiO$_7$ was also measured. All SEQUOIA data were collected at 5~K with an incident energy of 80 meV using the fine Fermi chopper. The $T_0$ frequency, Fermi chopper frequency, and energy resolution at the elastic line (full-width half-maximum) were 90~Hz, 480~Hz, and 2.15~meV. 

Lower-incident-energy neutron powder spectroscopy data were measured with the direct-geometry time-of-flight instrument CNCS \cite{Ehlers_2016} at the SNS using the same polycrystalline sample from the unpolarized HB-2A experiment. The CNCS data were collected using incident energies of $E_i =$1.55~meV  and 2.49~meV in the “high flux” chopper setting mode, which produced energy resolutions of 0.04~meV and 0.06~meV (full-width half-maximum) at the elastic line respectively. 

\section{Data Availability}

All the data supporting the findings of this study are available within the article and from the corresponding authors upon request. Source data are also provided as a Source Data file.

\bibliographystyle{naturemag}
\bibliography{references}

%\begin{thebibliography}{99} 
%\bibitem{}
%\end{thebibliography}

\section{Acknowledgments}

Research at the University of Tennessee is supported by the National Science Foundation, Division of Materials Research under Award No. NSF-DMR-2003117. The work at Michigan State University is supported by the U.S.DOE-BES under Contract DE-SC0023648. The work performed at NHMFL is supported by the NSF Cooperative Agreement No. DMR-1644779 and the State of Florida. A portion of this research used resources at the Spallation Neutron Source and the High Flux Isotope Reactor, which are DOE Office of Science User Facilities operated by Oak Ridge National Laboratory. The beam time was allocated to HB-2A (POWDER) on proposal number IPTS-29325.1 and IPTS-33066.1. Additional beam time was allocated to BL-1B (NOMAD) on proposal number IPTS-31286.1, BL-5 (CNCS) on proposal number IPTS-30936.1, and BL-17 (SEQUOIA) on IPTS-29415.1. A portion of the research at ORNL was supported by the DOE, Office of Science, Office of Advanced Scientific Computing Research (contract No. ERKJ387), and Office of Basic Energy Sciences (award No. KC0402020 under contract No. DE-AC05-00OR22725). Work by B.A.F. (pair distribution function analysis) was supported by the National Science Foundation LEAPS-MPS program through Grant No. NSF-DMR-2418438. G. Duan, R. Yu, N. Li, and X. F. Sun thank the support from the National Key R\&D Program of China (Grant No. 2023YFA1406500). G. Duan and R. Yu were also supported by the National Natural Science Foundation of China (Grant Nos. 12334008 and 12174441). N. Li and X. F. Sun  were also supported by the National Natural Science Foundation of China (Grant Nos. 12274388 and 12404043) and the Nature Science Foundation of Anhui Province (Grant No. 2408085QA024).

\section{Author Contributions}
A.B., H.D.Z., and A.A.A. conceived the project. A.B., H.D.Z., N.L., X.F.S., and E.S.C. synthesized the samples and performed the bulk characterization measurements. H.W. and W.X. collected and analyzed the single crystal x-ray diffraction data. A.B., Q.M., S.C., A.I.K., K.M.T., G.S. and A.A.A. participated in the neutron scattering data collection. A.B., Q.M., G.D., S.C., B.A.F., C.L., R.Y., H.D.Z. and A.A.A. contributed to the remaining data analysis. The manuscript was written by A.B. and A.A.A. Critical manuscript comments were provided by all authors. 

\section{Competing Interests}
The authors declare no competing interests.  

\end{document}

\title{Supplemental Information: Novel bipartite entanglement in the quantum dimer magnet Yb$_2$Be$_2$SiO$_7$}

\author{A. Brassington} 
\affiliation{Department of Physics and Astronomy, University of Tennessee, Knoxville, TN 37996, USA}

\author{Q. Ma} 
\affiliation{Neutron Scattering Division, Oak Ridge National Laboratory, Oak Ridge, TN 37831, USA}

\author{G. Duan} 
\affiliation {School of Physics and Beijing Key Laboratory of Optoelectronic Functional Materials and Micro-nano Devices, Renmin University of China, Beijing 100872, China}

\author{S. Calder} 
\affiliation{Neutron Scattering Division, Oak Ridge National Laboratory, Oak Ridge, TN 37831, USA}

\author{A.I. Kolesnikov} 
\affiliation{Neutron Scattering Division, Oak Ridge National Laboratory, Oak Ridge, TN 37831, USA}

\author{K.M. Taddei} 
\affiliation{Neutron Scattering Division, Oak Ridge National Laboratory, Oak Ridge, TN 37831, USA}

\author{G. Sala} 
\affiliation{Oak Ridge National Laboratory, Oak Ridge, TN 37831, USA}

\author{E.S. Choi} 
\affiliation{National High Magnetic Field Laboratory and Department of Physics, Florida State University, Tallahassee, Florida 32310, USA}

\author{H. Wang} 
\affiliation {Department of Chemistry, Michigan State University, East Lansing, Michigan 48824, United States}

\author{W. Xie} 
\affiliation {Department of Chemistry, Michigan State University, East Lansing, Michigan 48824, United States}

\author{B.A. Frandsen}
\affiliation{Department of Physics and Astronomy, Brigham Young University, Provo, UT 84602, USA}

\author{N. Li}
\affiliation{Anhui Provincial Key Laboratory of Magnetic Functional Materials and Devices, Institutes of Physical Science and Information Technology, Anhui University, Hefei, Anhui 230601, People's Republic of China}

\author{X.F. Sun}
\affiliation{Anhui Provincial Key Laboratory of Magnetic Functional Materials and Devices, Institutes of Physical Science and Information Technology, Anhui University, Hefei, Anhui 230601, People's Republic of China}

\author{C. Liu}
\affiliation {School of Engineering, Dali University, Dali, Yunnan 671003, China}

\author{R. Yu} 
\affiliation {School of Physics and Beijing Key Laboratory of Optoelectronic Functional Materials and Micro-nano Devices, Renmin University of China, Beijing 100872, China}
\affiliation {Key Laboratory of Quantum State Construction and Manipulation (Ministry of Education), Renmin University of China, Beijing 100872, China}

\author{H.D. Zhou} \altaffiliation{\href{mailto:hzhou10@utk.edu}{hzhou10@utk.edu}}
\affiliation{Department of Physics and Astronomy, University of Tennessee, Knoxville, TN 37996, USA}

\author{A.A. Aczel} \altaffiliation{\href{mailto:aczelaa@ornl.gov}{aczelaa@ornl.gov}}
\affiliation{Neutron Scattering Division, Oak Ridge National Laboratory, Oak Ridge, TN 37831, USA}
 
\maketitle 

\date{\today}

\section{Structural Characterization}

The phase purity of the Yb$_2$Be$_2$SiO$_7$ polycrystalline and single crystal samples was confirmed via room-temperature powder x-ray diffraction (XRD) using a HUBER imaging plate Guinier camera 670 with Cu radiation ($\lambda =$~1.54059~\AA). The single crystal samples were first ground into a fine powder for this measurement. The XRD refinements were performed with the software package FULLPROF \cite{RODRIGUEZCARVAJAL_1993} using the structure of Y$_2$Be$_2$SiO$_7$ \cite{Kuzmicheva_2002} as a starting reference.

As an additional check, single crystal XRD measurements were carried out on a Bruker Eco Quest X-ray Diffractometer using Mo radiation ($\lambda =$~0.71073~\AA) at 300 K and the structure refinement was completed using the Bruker SHELXTL Software Package. The main results are presented in Supplementary Tables~\ref{table:SXRD1} and \ref{table:SXRD2} and agree well with the previous powder refinement \cite{Brassington_2024}. 

\begin{table}
\caption{Single crystal refinement parameters for Yb$_2$Be$_2$SiO$_7$ at 300 K}
\begin{tabular}{c c } 
\hline 
Formula weight & 504.19 g/mol \\
Space Group & $P\bar{4}2_1 m$ \\
Unit cell & a = 7.207(1) \AA \\
Unit cell & c = 4.719(1) \AA \\
Volume & 245.1(2) \AA$^3$ \\
Density (calculated) & 6.832 g/cm$^3$ \\
Extinction coefficient & 0.082(3) \\
Absorption coefficient & 38.147 mm$^{-1}$ \\
F(000) & 436 \\
2$\theta$ range & 8.00 to 66.10$^{\circ}$ \\
Total Reflections & 4477 \\
Independent reflections & 513 [Rint = 0.0550] \\
Refinement method & Full-matrix least-squares F$^2$ \\
Absolute structure parameter & 0.03(4) \\
Data / restraints / parameters & 559 / 0 / 21 \\
Final R indices & R$_1$(I$\>$2$\sigma$(I)) = 0.0189; \\
                & wR$_2$ (I$\>$2$\sigma$(I)) = 0.0440 \\
                & R$_1$ (all) = 0.0192; \\ 
                & wR$_2$ (all) = 0.0442 \\
Largest diff. peak and hole & +1.972 e/\AA$^{-3}$, -1.249 e/\AA$^{-3}$\\
R.M.S. deviation from mean & 0.370 e/\AA$^{-3}$ \\
Goodness-of-fit on F$^{2}$ & 1.149  \\
\hline 
\end{tabular}
\label{table:SXRD1} 
\end{table}

\begin{table}
\caption{Atomic coordinates and isotropic displacement parameters $U_{eq}$ (\AA$^2$) from single-crystal XRD}
\begin{tabular}{c c c c c c} 
\hline
Atom   & Wyck. & x \hspace{1 pt} & \hspace{1 pt} y \hspace{1 pt} & \hspace{1 pt} z \hspace{1 pt} & \hspace{1 pt}$U_{eq}$ \hspace{1 pt}\\
\hline
Yb    & 4e & 0.1589(1) & 0.6589(1) & 0.50772(1)  & 0.0040(1)\\

Be    & 4e & 0.637(1) & 0.137(1) & 0.960(2)  & 0.006(2)\\

Si    & 2a & 0 & 0 & 0  & 0.0033(4)\\

O$_1$ & 2c & 0 & 1/2 & 0.185(2) & 0.005(1)\\

O$_2$ & 4e & 0.6423(6) & 0.1423(6) & 0.295(1)  & 0.005(1)\\

O$_3$ & 8f & 0.0824(7) & 0.1624(6) & 0.203(1)  & 0.0058(8)\\
\hline

\end{tabular}
\label{table:SXRD2} 
\end{table}

Neutron powder diffraction was performed using the high-resolution HB-2A powder diffractometer at the High Flux Isotope Reactor of Oak Ridge National Laboratory. The experimental details are provided in the main article. The 0.25~K and 2~K diffraction patterns are well-described by the known room-temperature crystal structure. The refined lattice parameters at 0.25~K are $a = b = 7.2155(2)$~\AA~and $c = 4.7171(2)$~\AA, which implies that there is some sample-dependence particularly in the basal-plane values. The fractional coordinates at 0.25~K are presented in Supplementary Table~\ref{table:HB2A}. 

\begin{table}
\caption{Fractional coordinates for Yb$_2$Be$_2$SiO$_7$ from neutron powder diffraction at 0.25 K}
\begin{tabular}{c c c c c} 
\hline
Atom   & Wyck. & x \hspace{1 pt} & \hspace{1 pt} y \hspace{1 pt} & \hspace{1 pt} z \hspace{1 pt}\\
\hline
Yb    & 4e & 0.1590(3) & 0.6590(3) & 0.506(1)  \\

Be    & 4e & 0.6387(4) & 0.1387(4) & 0.955(1) \\

Si    & 2a & 0 & 0 & 0 \\

O$_1$ & 2c & 0 & 1/2 & 0.181(2) \\

O$_2$ & 4e & 0.6405(6) & 0.1405(6) & 0.303(1) \\

O$_3$ & 8f & 0.0846(5) & 0.1658(6) & 0.201(1)  \\
\hline

\end{tabular}
\label{table:HB2A} 
\end{table}

Time-of-flight neutron powder diffraction data were collected at 100~K using the NOMAD beam line at the Spallation Neutron Source on three samples: polycrystalline Yb$_2$Be$_2$SiO$_7$, finely ground single crystals of Yb$_2$Be$_2$SiO$_7$, and polycrystalline Er$_2$Be$_2$SiO$_7$. The goal of this experiment was to identify evidence for local structural distortions that could be responsible for the Yb$^{3+}$ crystal field broadening observed in the SEQUOIA data presented in the main text. The NOMAD data reduction was completed with the Advanced Diffraction Environment (ADDIE) suite \cite{Mcdonnell_2017}. Pair Distribution Function (PDF) refinements were completed using the software package PDFGUI \cite{Farrow_2007} with a maximum $Q$ value of 30~\AA$^{-1}$. The pair distribution function, $G(r)$, as a function of distance, r(\AA), is shown for the three samples in Supplementary Fig.~\ref{figS1}. All three PDF patterns are well-explained by the known crystal structure of these materials with no clear evidence for significant deviations that would arise from local structure distortions. It is also worth noting that the PDF patterns for polycrystalline and crushed single crystal Yb$_2$Be$_2$SiO$_7$ are nearly identical.

\begin{figure}
\scalebox{0.24}{\includegraphics{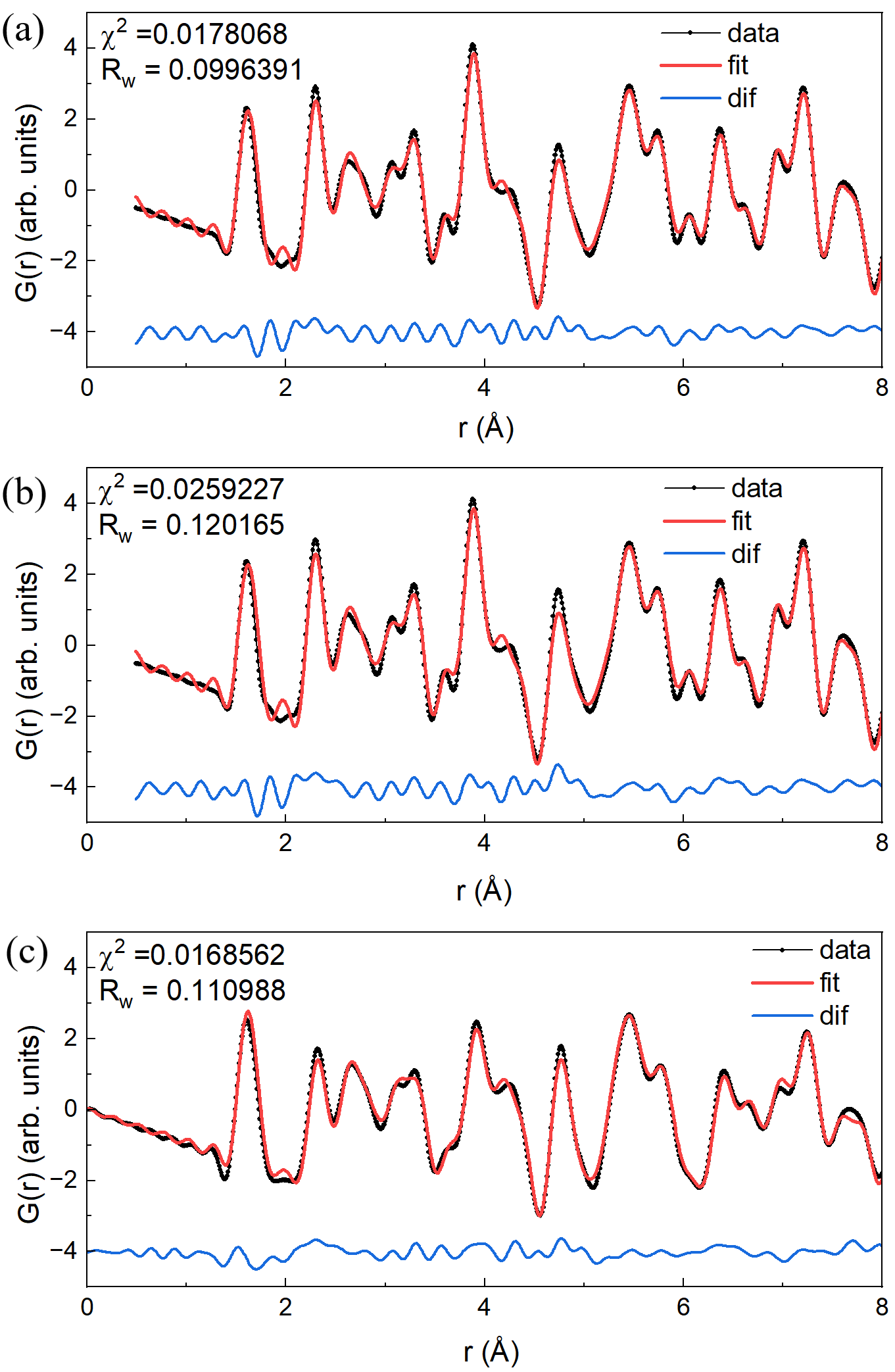}}
\caption{{\bf Local structure investigation.} The pair distribution function $G(r)$ vs distance for (a) polycrystalline Yb$_2$Be$_2$SiO$_7$, (b) crushed single crystal Yb$_2$Be$_2$SiO$_7$, and (c) polycrystalline Er$_2$Be$_2$SiO$_7$. The data are described well by the known crystal structures for these materials, so there is no evidence for local structure distortions.}
\label{figS1}
\end{figure}

\section{Half-Polarized Neutron Powder Diffraction}

The magnetic susceptibility $\chi$ is a measure of the magnetization of a material in an applied magnetic field. In most crystalline materials, the influence of spin-orbit coupling induces anisotropy that results in $\chi$ taking the form of a second-rank tensor rather than a scalar quantity. The components of this tensor can be determined by the half-polarized neutron powder diffraction (pNPD) technique in the linear $M/H$ regime and they provide insight into the local anisotropy of the magnetic ion \cite{Baral_2025}. The atomic site symmetry is used to establish appropriate constraints for $\chi$. Spin-up and spin-down neutron diffraction patterns with intensities of I$^+$ and I$^-$ are measured separately, and then Rietveld refinements of the sum and difference patterns (given by I$^+$ + I$^-$ and I$^+$ - I$^-$ respectively) are performed using the software CrysPy \cite{Kibalin_2019}. 

The results of these Rietveld refinements for the Yb$_2$Be$_2$SiO$_7$ HB-2A data collected at 5~K and 1.5~T are shown in Supplementary Fig.~\ref{figS2}. The local site susceptibility tensor extracted from this analysis is given by:
$$ \chi_{ij}  = \begin{pmatrix}
-0.12(3) & -0.02(2) & 0.05(4)\\
-0.02(2) & -0.12(3) & 0.05(4)\\
0.05(4)  & 0.05(4) & 2.36(3)
\end{pmatrix} $$

The magnetization ellipsoids' principal axis directions and magnitudes can be obtained from the local site susceptibility tensor. For Yb$_2$Be$_2$SiO$_7$, we find that the two principal $g$-tensor directions not constrained by symmetry are nearly aligned with the crystallographic $[001]$-axis and the dimer bond direction ($[110]$ and $[1\bar{1}0]$ for the two dimer sublattices). The magnitudes of the principal axes are 2.32 $\mu_B$/T, 0.14 $\mu_B$/T, and 0.10 $\mu_B$/T, with the largest value for the pseudo-$[001]$-axis direction. The magnetization ellipsoids are plotted in Fig.~2(e) of the main manuscript. 

\begin{figure}
\scalebox{0.235}{\includegraphics{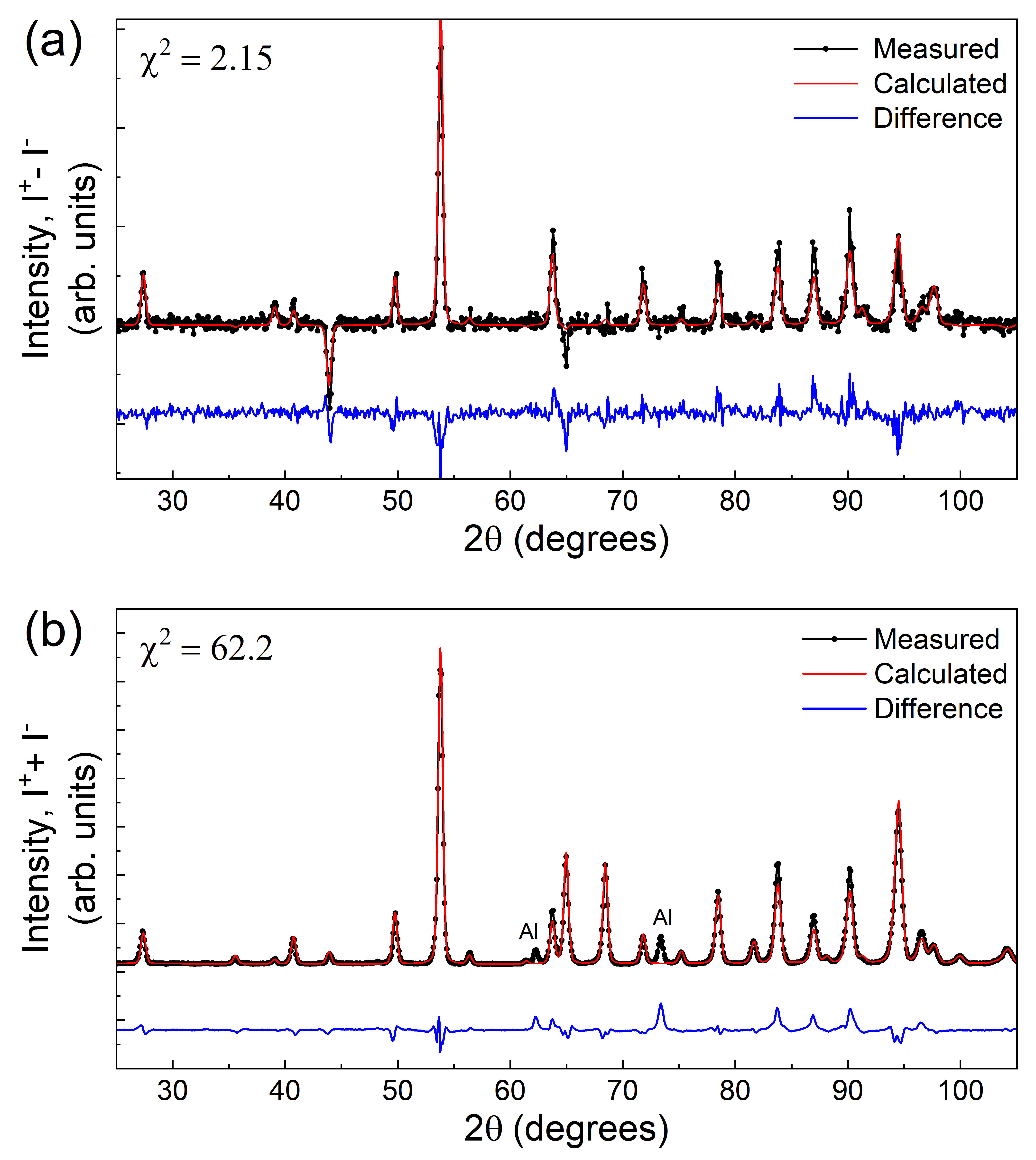}}
\caption{{\bf pNPD Rietveld refinement results.} (a) The difference pattern, I$^+$ - I$^-$, and (b) the sum pattern, I$^+$ + I$^-$, obtained from the pNPD measurements at 5 K under an applied field of 1.5 T. For each case, the Rietveld refinement is superimposed on the data and the fit residual is shown below it. The two peaks from the Al sample can are labeled in (b).}
\label{figS2}
\end{figure}

\section{Isolated Dimer Model}

For Yb$_2$Be$_2$SiO$_7$, the first excited CEF level is well-separated from the ground state which ensures that an effective spin-1/2 model is applicable at sufficiently low-temperatures. We construct the spin operators within a dimer unit as:
\begin{eqnarray}
S^{\alpha}_{m} &=& \dfrac{1}{2} (\sigma_{\alpha} \otimes \mathbf{1})\nonumber \\
S^{\beta}_{n} &=& \dfrac{1}{2} (\mathbf{1} \otimes \sigma_{\beta})
\label{spins}
\end{eqnarray}
where $\alpha,\beta$ = $x,y,z$, $\mathbf{1}$ represents the identity matrix, $\sigma_{\alpha, \beta}$ denotes the Pauli matrices and $\otimes$ is the Kronecker product.

The strong spin-orbit coupling of the Yb$^{3+}$ ions and the low-symmetry of the crystal structure ensures that a Hamiltonian with XYZ intradimer exchange is required to explain the data. The proposed intradimer Hamiltonian with the associated Zeeman term is shown in Eq.~1 of the main manuscript. Diagonalizing this Hamiltonian generates the following eigenvalues: 
\begin{align}
E_0 = \frac{1}{4}(-J^{A/B}_{xx}+J^{A/B}_{yy}+J^{A/B}_{zz}) \nonumber \\
E_1 = \frac{1}{4}(J^{A/B}_{xx}+J^{A/B}_{yy}-J^{A/B}_{zz}) \nonumber \\
E_2 = \frac{1}{4}(J^{A/B}_{xx}-J^{A/B}_{yy}+J^{A/B}_{zz}) \nonumber \\
E_3 = \frac{1}{4}(-J^{A/B}_{xx}-J^{A/B}_{yy}-J^{A/B}_{zz})
\end{align}

\begin{figure}
\scalebox{0.31}{\includegraphics{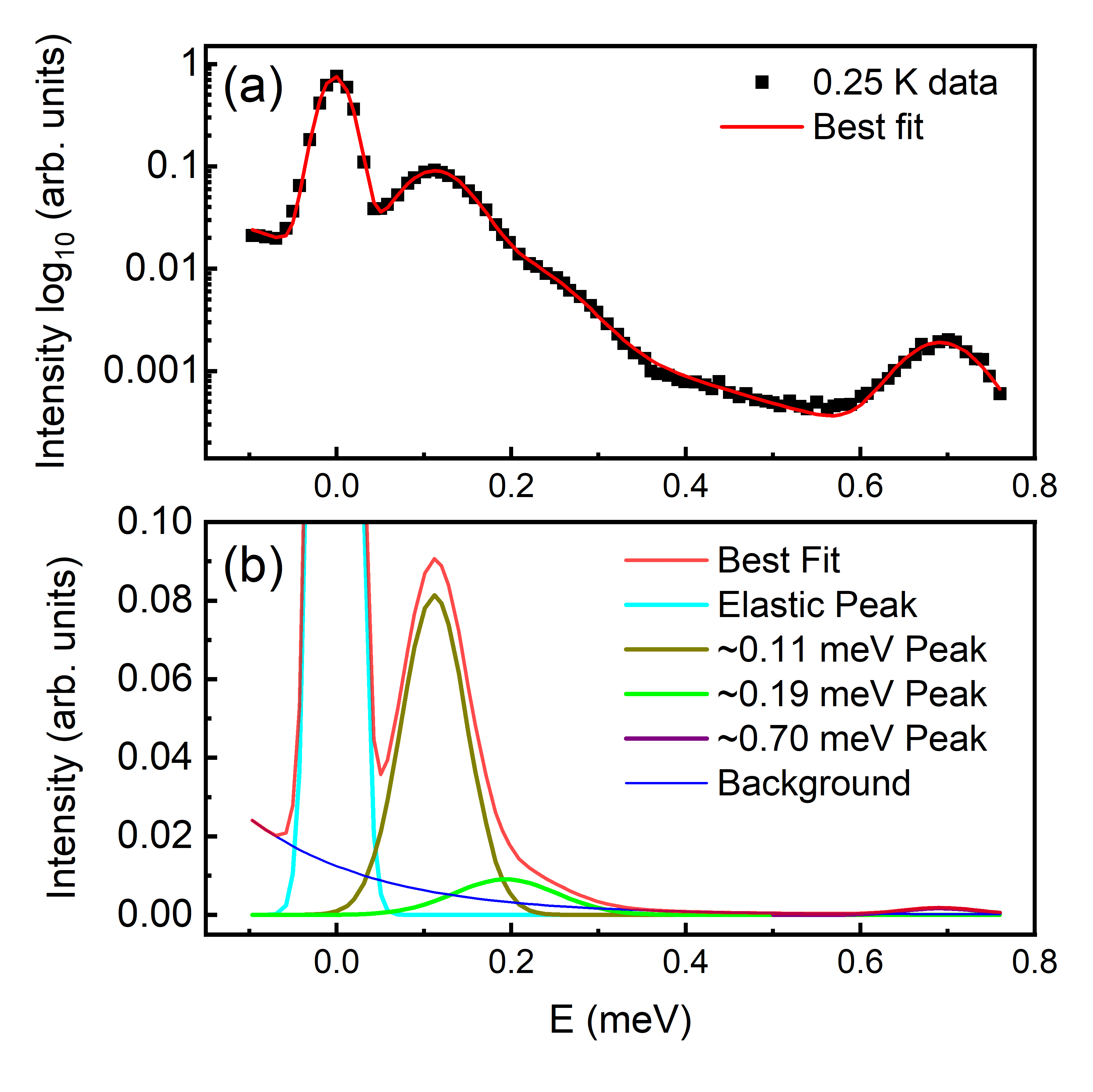}}
\caption{{\bf Low-energy neutron spectroscopy peak fitting.} (a) Constant-$Q$ cut of CNCS data ($Q$-integration range [0.3, 1.8]~\AA$^{-1}$) with $E_i =$~1.55~meV and $T =$~0.25~K. The best fitting result using a function with four Gaussian peaks, a decaying exponential term, and a constant is superimposed on the data. (b) The contribution of each Gaussian peak to the final fit result. The extracted peak parameters are given in Supplementary Table S4.}
\label{figS3}
\end{figure}

\begin{figure}

\scalebox{0.23}{\includegraphics{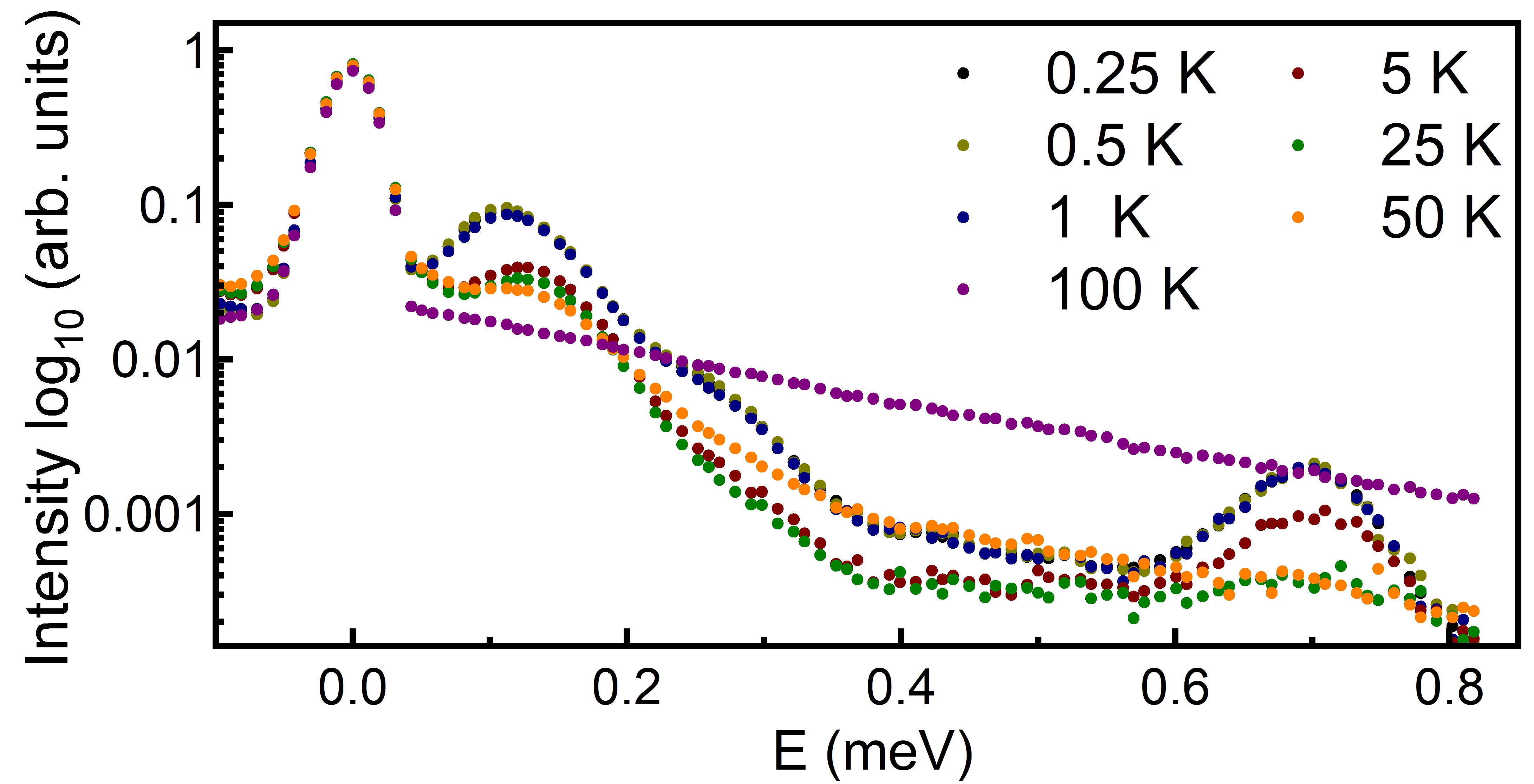}}
\caption{{\bf $T$-dependence of low-energy excitations.} (a) Constant-$Q$ cut of CNCS data ($Q$-integration range [0.3, 1.8]~\AA$^{-1}$) with $E_i =$~1.55~meV at several temperatures between 0.25 K and 100 K.}
\label{figS4}
\end{figure}

\begin{table}
\caption{CNCS data fitting results at 0.25~K with E$_i$ = 1.55 meV using a four Gaussian peak model}
\begin{tabular}{|c | c | c | c | c | c|} 
\hline
Peak center & Peak width(FWHM)  & Peak Amplitude \\
\hline
-0.0019(2)    & 0.0390(3) & 1   \\
\hline
0.111(2)    & 0.088(4) & 0.25(3)
  \\
\hline
0.19(4)    & 0.14(5) & 0.04(3)
  \\
\hline
0.692(8)    & 0.10(2) & 0.006(2)
  \\
\hline

\end{tabular}
\label{table:CNCS-fits} 
\end{table}

\begin{figure}
\scalebox{0.145}{\includegraphics{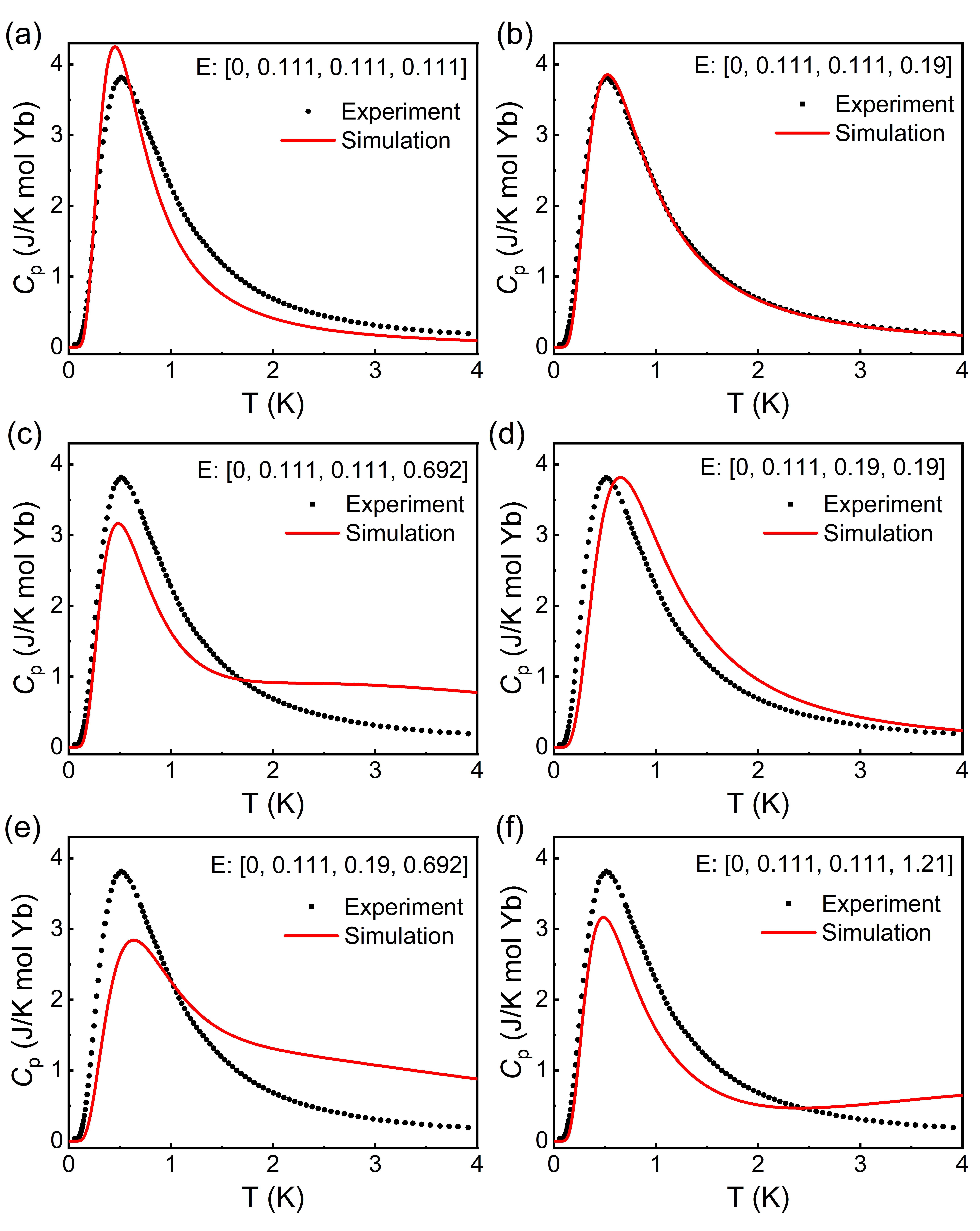}}
\caption{{\bf Zero-field heat capacity modeling.} Simulations for different single-dimer models are superimposed on the data in panels (a-f). The simulation with a doubly-degenerate 0.11~meV excitation and a non-degenerate 0.19~meV excitation describes the data best.}
\label{figS5}
\end{figure} 

We used the zero-field heat capacity and the neutron spectroscopy data to establish the experimental eigenvalues for Yb$_2$Be$_2$SiO$_7$. Neutron spectroscopy data collected at CNCS with $E_i =$~1.55 meV reveals three magnetic excitations with full-width half-maximum values only slightly greater than the expected instrument resolution. The precise energy and peak width values were extracted from the constant-$Q$ cut shown in Supplementary Fig.~\ref{figS3} using a fitting function consisting of four Gaussian peaks, a decaying exponential term, and a constant. These parameters are provided in Supplementary Table S4. The temperature-dependence of these modes, presented in Supplementary Fig.~\ref{figS4}, suggests that they have a magnetic origin. There is also a fourth higher-energy magnetic excitation that was observed in the $E_i =$~2.49 meV data presented in the main manuscript. To assess which excitations could be associated with single dimer physics, we simulated our low-$T$ heat capacity data with the function \cite{Probert_2012}:
\begin{multline}
C_{m}(T) = \frac{1}{k_B T^2} \Bigl\{-\left(\frac{1}{Z} \sum_{j} E_{j} ~exp\left(-\frac{E_{j}}{k_B T}\right)\right)^{2} \\+ \frac{1}{Z} \sum_{j} E^{2}_{j} ~exp\left(-\frac{E_{j}}{k_B T}\right) \Bigl\}
\end{multline}
where $E_j$ is the eigenvalue of dimer state $j$, $Z = \sum_{j} exp(-\frac{E_{j}}{k_B T})$ is the partition function, and $k_B$ is the Boltzmann factor. The XYZ Hamiltonian can yield up to three single-dimer excitations, so we tried different eigenvalue combinations in our simulations that were consistent with the four energy levels observed in neutron spectroscopy. Several of these simulation results are presented in Supplementary Fig.~\ref{figS5}. We found that the two higher-energy excitations have negligible contributions to the measured heat capacity, so we do not consider them further in our single-dimer models. The best agreement between the data and the simulation is found for a single-dimer model with a doubly-degenerate 0.11~meV excitation and a non-degenerate 0.19~meV excitation. 

\begin{figure*}
\scalebox{0.24}{\includegraphics{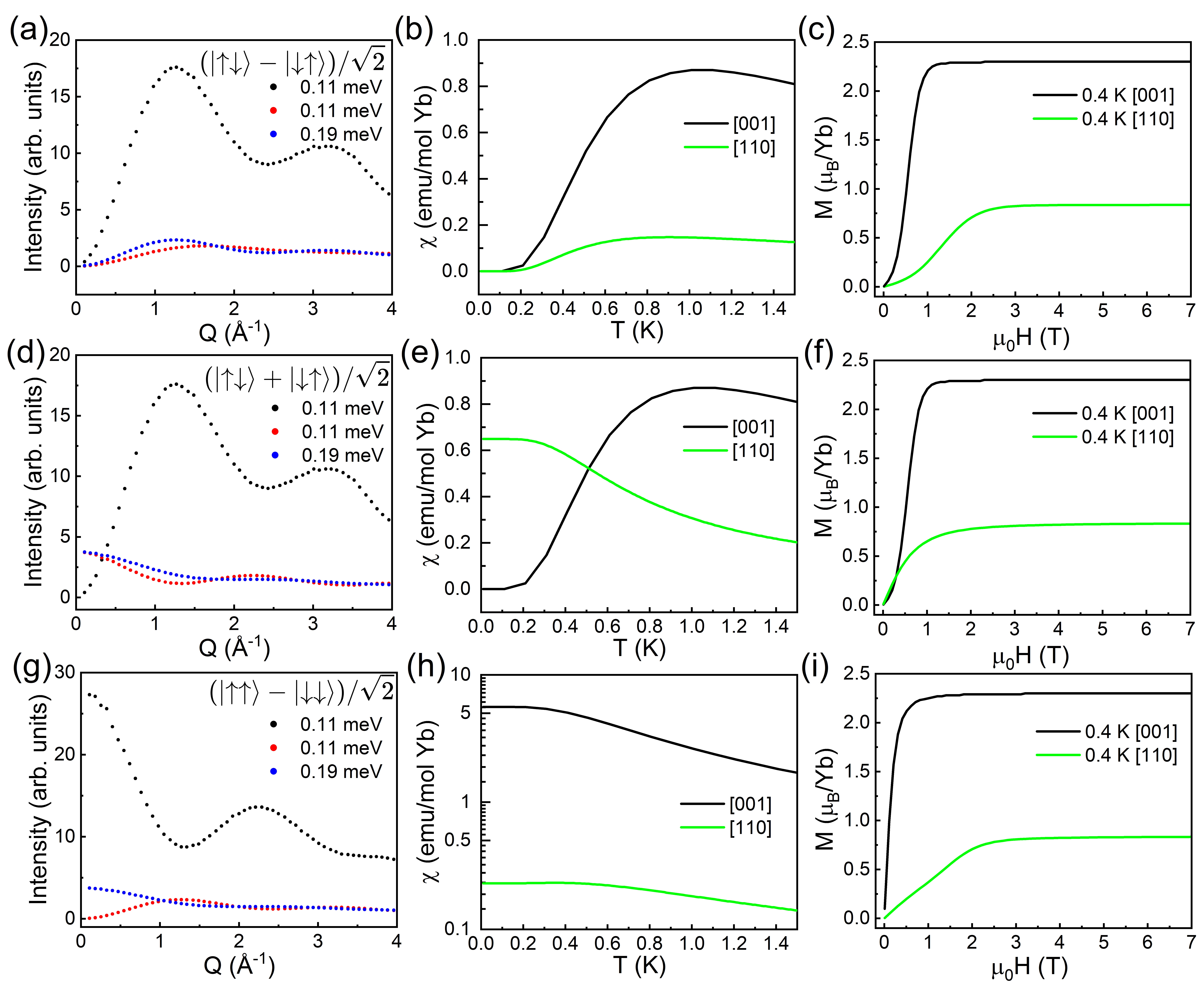}}
\caption{{\bf Simulations for XYZ dimer models with different ground states.} (a) Simulated powder dynamical structure factor $S(Q)$ for the three excitations, (b) simulated single crystal magnetic susceptibility as a function of temperature, and (c) simulated single crystal anisotropic magnetization at 0.4~K for an isolated dimer model with anisotropic exchange using the parameters $J_{xx} = 0.03$~meV, $J_{yy} = 0.19$~meV, and $J_{zz} = 0.19$~meV with the dimer ground state $\frac{1}{\sqrt{2}}({\ket{\uparrow \downarrow} - \ket{\downarrow \uparrow}})$. (d-f) Similar simulations for a second isolated dimer model using the parameters $J_{xx} = -0.19$~meV, $J_{yy} = -0.03$~meV, and $J_{zz} = 0.19$~meV with the dimer ground state $\frac{1}{\sqrt{2}}({\ket{\uparrow \downarrow} + \ket{\downarrow \uparrow}})$. (g-i) Similar simulations for a third isolated dimer model using the parameters $J_{xx} = 0.19$~meV, $J_{yy} = -0.03$~meV, and $J_{zz} = -0.19$~meV with the dimer ground state $\frac{1}{\sqrt{2}}({\ket{\uparrow \uparrow} - \ket{\downarrow \downarrow}})$.} 
\label{figS6}
\end{figure*} 

The XYZ Hamiltonian has 12 solutions that are consistent with this established eigenvalue scheme, so we simulated the neutron spectroscopy, magnetization, magnetic susceptibility, and field-induced heat capacity for each of these models. The neutron spectroscopy simulations of the dynamical structure factor for transitions between the single dimer levels $j$ and $k$ of the XYZ model were performed using the expression \cite{Andres_1999}:
\begin{multline}
S(\vec{Q},E) = A \sum_{j,k} \exp\left(- \frac{E_j}{k_B T}\right) \sum_{\alpha,\beta} \left(\delta_{\alpha\beta} - \frac{Q_{\alpha} Q_{\beta}}{Q^{2}}\right)\times \\ \sum_{m,n} f^{*}_{n}(Q) f_{m}(Q) ~\exp(i\vec{Q}\cdot(\vec{R}_m-\vec{R}_n)) \times \\ \bra{\psi_{j}} M^{\alpha}_{m} \ket{\psi_{k}} \times \bra{\psi_{k}} M^{\beta}_{n} \ket{\psi_{j}} \delta(E + E_{j} - E_{k})
\end{multline}
where $\alpha, \beta = {x,y,z}$, $S_m^\alpha$ refers to the $\alpha$ component of the effective spin-1/2 operator for magnetic ion $m$, $g^{m}_{\alpha\beta}$ represents one component of the $g$-tensor for magnetic ion $m$, $\vec{R}_m$ and $f_m(Q)$ are the position vector and magnetic form factor for magnetic ion $m$, $\psi_{j}$ is the eigenfunction of dimer state $j$, and $M^{\alpha}_{m} = \sum_\gamma g^{m}_{\alpha\gamma} S^{\gamma}_{m}$. The quantity $A$ includes a constant and the Debye-Waller factor $exp(-2W)$. The neutron spectroscopy simulation was powder-averaged to facilitate direct comparison to the CNCS data using the equation:
\begin{equation}
S(Q,E) = \int \frac{d\Omega}{4\pi}S(\vec{Q},E)
\end{equation}

%\begin{multline}
%S(\vec{Q},E) = A \sum_{j,k} exp\left(- \frac{E_j}{k_B T}\right) \sum_{\alpha,\beta} \left(\delta_{\alpha,\beta} - %\frac{Q_{\alpha} Q_{\beta}}{Q^{2}}\right)\times \\ \sum_{m,n} f^{*}_{n}(Q) f_{m}(Q) ~exp(i\vec{Q}\cdot(\vec{R}_m-\vec{R}_n)) %\times \\ \bra{\psi_{j}} g^{m}_{\alpha\beta} S^{\alpha}_{m} \ket{\psi_{k}} \times \bra{\psi_{k}} g^{n}_{\alpha\beta} %S^{\beta}_{n} \ket{\psi_{j}}  \delta(E + E_{j} - E_{k})
%\end{multline}

\begin{figure*}
\scalebox{0.4}{\includegraphics{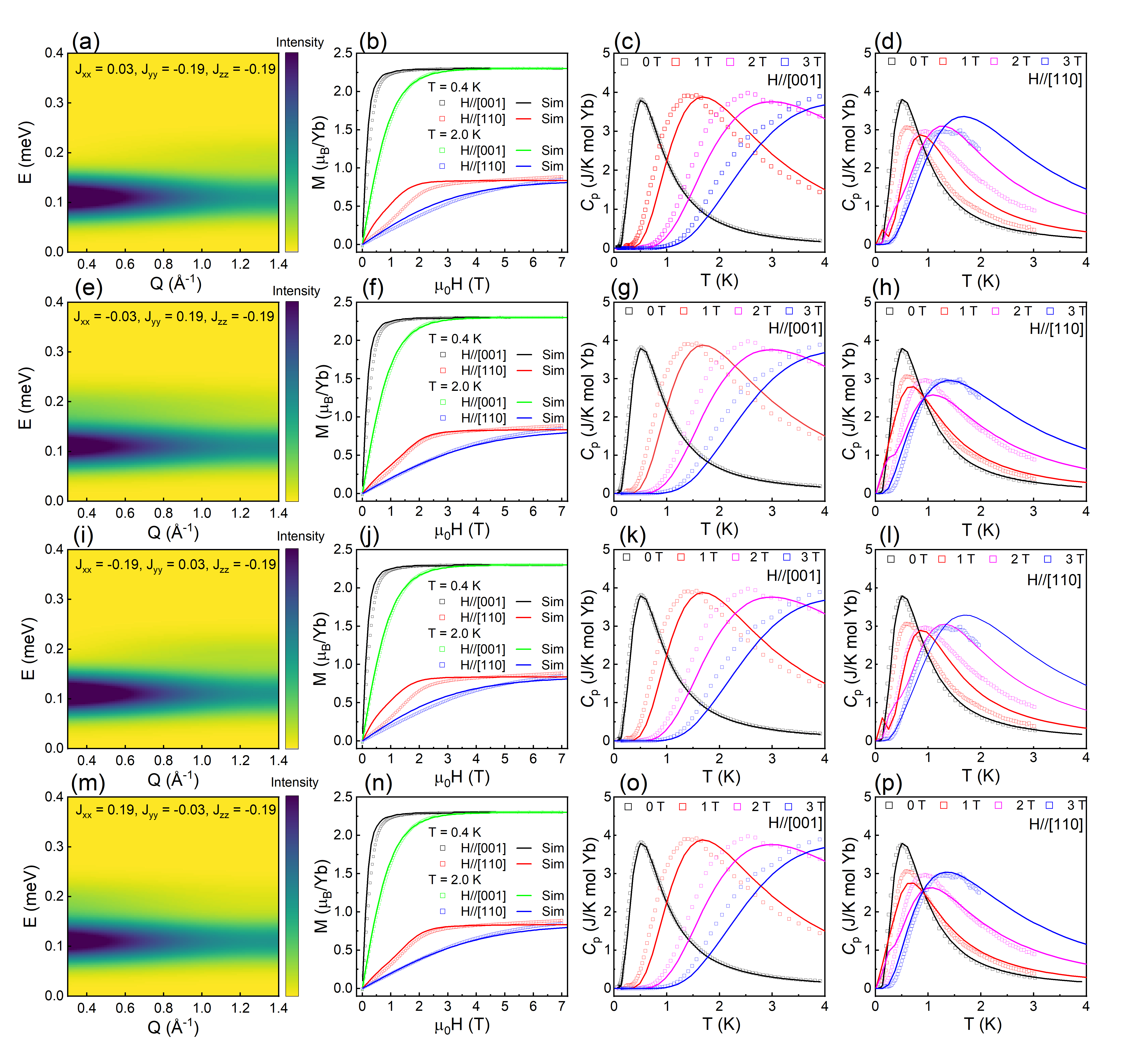}}
\caption{{\bf Detailed comparison of the four most probable XYZ dimer models for Yb$_2$Be$_2$SiO$_7$.} (a) Simulated dynamical structure factor $S(Q,E)$ for the three excitations, (b) simulated and measured anisotropic magnetization at both 0.4~K and 2~K, (c) simulated and measured heat capacity data for $\vec{H} \parallel [001]$, and (d) simulated and measured heat capacity data for $\vec{H} \parallel [110]$. The simulations correspond to an isolated dimer model with anisotropic exchange using the parameters $J_{xx} = 0.03$~meV, $J_{yy} = -0.19$~meV, and $J_{zz} = -0.19$~meV. Similar dynamical structure factor simulations and the same bulk characterization data with similar simulations superimposed on it for (e-h) an isolated dimer model with $J_{xx} = -0.03$~meV, $J_{yy} = 0.19$~meV, and $J_{zz} = -0.19$~meV, (i-l) an isolated dimer model with $J_{xx} = -0.19$~meV, $J_{yy} = 0.03$~meV, and $J_{zz} = -0.19$~meV, and (m-p) an isolated dimer model with $J_{xx} = 0.19$~meV, $J_{yy} = -0.03$~meV, and $J_{zz} = -0.19$~meV. The second and fourth set of simulations show the best agreement with the experimental data.}
\label{figS7}
\end{figure*}

The magnetization simulations were performed using the expression: 
\begin{multline}
M(\vec{H},T) = \frac{1}{N}\sum_{m} \frac{1}{Z} \sum_{j} \bra{\psi_{j}(\vec{H})} \mu_B~\vec{g}_m \cdot \vec{S}_m \ket{\psi_{j}(\vec{H})} \times \\ exp\left(-\frac{E_{j}}{k_B T}\right) 
\end{multline}
where $N$ is the number of magnetic ions in a unit cell, $\psi_j (\vec{H})$ is the eigenfunction of dimer state $j$ in an applied magnetic field $\vec{H}$, $\vec{g}_m$ is the $g$-tensor of the $m$th spin, and $\vec{S}_m$ is the effective spin-1/2 operator of the $m$th spin. The DC susceptibility was simulated by taking derivatives of the magnetization as:
\begin{equation}
\chi(\vec{H},T) = \frac{\partial M(\vec{H},T)}{\partial H}.
\end{equation}
Within a linear response regime of small $\vec{H}$, $\chi(\vec{H},T)$ remains a constant. The field-dependent heat capacity simulations were performed using Supplementary Eq.~S3 with modified eigenvalues obtained by diagonalizing the XYZ Hamiltonian with the appropriate Zeeman term included. 

Supplementary Fig.~\ref{figS6} presents simulations of the powder-averaged dynamical structure factors, the low-$T$ magnetic susceptibility, and the $\vec{H} \parallel$~[001] and [110] magnetization for representative XYZ dimer models with the known $g$-tensor and eigenvalues for Yb$_2$Be$_2$SiO$_7$ and $\frac{1}{\sqrt{2}}({\ket{\uparrow \downarrow} - \ket{\downarrow \uparrow}})$, $\frac{1}{\sqrt{2}}({\ket{\uparrow \downarrow} + \ket{\downarrow \uparrow}})$, and $\frac{1}{\sqrt{2}}({\ket{\uparrow \uparrow} - \ket{\downarrow \downarrow}})$ ground states. In the regime $Q \le$~1.6~\AA$^{-1}$, the $S_z =$~0 models have intense modes with a dynamical structure factor, 
\begin{equation}
S(Q) = A\left(1-\frac{sin(Qd)}{Qd}\right)
\end{equation}
where $A$ is a constant and $d$ is the intradimer distance. One can distinguish between them by measuring the $Q$-dependence of the weaker mode intensities, the low-$T$ magnetic susceptibility, or the low-$T$ anisotropic magnetization. All models with $\frac{1}{\sqrt{2}}({\ket{\uparrow \downarrow} - \ket{\downarrow \uparrow}})$ ground states have low-field magnetization plateaus and sharp drops in the magnetic susceptibility with decreasing $T$ when the magnetic field is applied both parallel and perpendicular to the quantization axis, while the models with $\frac{1}{\sqrt{2}}({\ket{\uparrow \downarrow} + \ket{\downarrow \uparrow}})$ ground states only have low-field magnetization plateaus and sharp drops in the magnetic susceptibility with decreasing $T$ when the magnetic field is applied along the quantization axis. All models with $\frac{1}{\sqrt{2}}({\ket{\uparrow \uparrow} + \ket{\downarrow \downarrow}})$ and $\frac{1}{\sqrt{2}}({\ket{\uparrow \uparrow} - \ket{\downarrow \downarrow}})$ ground states have intense modes with a dynamical structure factor well-described by the function 
\begin{equation}
S(Q) = A\frac{sin(Qd)}{Qd}
\end{equation}
in the regime $Q \le$~1.6~\AA$^{-1}$ and exhibit no sharp drops in the susceptibility with decreasing $T$ or low-field magnetization plateaus regardless of the applied field direction.

We identified four models out of the 12 possibilities that were consistent with the $Q$-dependence and energy of the most intense mode measured by neutron spectroscopy.  The eigenvectors for these four models are as follows:
\begin{align}
\ket{\psi_0} = \frac{1}{\sqrt{2}}({\ket{\uparrow \uparrow} - \ket{\downarrow \downarrow}}) \nonumber \\
\ket{\psi_1} = \frac{1}{\sqrt{2}}({\ket{\uparrow \downarrow} + \ket{\downarrow \uparrow}}) \nonumber \\
\ket{\psi_2} = \frac{1}{\sqrt{2}}({\ket{\uparrow \uparrow} + \ket{\downarrow \downarrow}}) \nonumber \\
\ket{\psi_3} = \frac{1}{\sqrt{2}}({\ket{\uparrow \downarrow} - \ket{\downarrow \uparrow}}) 
\end{align}

The simulation results of the four models are presented in Supplementary Fig.~\ref{figS7}.

The dimer ground state for all four models is characterized by $S_z \neq 0$ wavefunctions, which are stabilized by the large ferromagnetic $J_{zz} = -0.19$~meV that is common to all of them. The four models only produce two different sets of bulk characterization simulations, as exchanging $J_{xx}$ and $J_{yy}$ only affects the $Q$-dependent intensity of the weak modes measured by neutron spectroscopy. One set of models produces simulations that show much better agreement with the bulk characterization data. Although the $Q$-dependence of the weak mode intensities is not exactly the same for these two models, our neutron powder spectroscopy data is not sufficient for differentiating between them conclusively. The exchange parameters for one of these models are $J_{xx} =~0.19$~meV, $J_{yy} = -0.03$~meV, and $J_{zz} = -0.19$~meV and it has an entangled dimer ground state of $\frac{1}{\sqrt{2}}({\ket{\uparrow \uparrow} - \ket{\downarrow \downarrow}})$. The second model has exchange parameters of $J_{xx} =~-0.03$~meV, $J_{yy} = 0.19$~meV, and $J_{zz} = -0.19$~meV and it has an entangled dimer ground state of $\frac{1}{\sqrt{2}}({\ket{\uparrow \uparrow} + \ket{\downarrow \downarrow}})$.

\bibliographystyle{naturemag}
\bibliography{references}